\documentclass[twocolumn]{aastex631}

\usepackage{graphicx}	
\usepackage{amsmath}	
\usepackage{amssymb}	
\usepackage{xcolor}
\usepackage{natbib}
\usepackage{multirow}
\usepackage{soul}

\def\omA{\omega_{\rm A}}

\newcommand{\om}{2\Omega}

\def\bu{\boldsymbol{u}}
\def\bB{\boldsymbol{B}}
\def\bb{\boldsymbol{b}}

\def\tnabla{\widetilde{\nabla}}
\def\tbB{\widetilde{\bB}}
\def\tbu{\widetilde{\bu}}
\def\tvart{\widetilde{\vartheta}}

\def\kTI{k_{\rm TI}}

\def\BV{Brunt-V\"ais\"al\"a}
\def\keta{k_{\eta}}
\def\kkappa{k_{\kappa}}
\def\knu{k_{\nu}}

\def\Eq{Equation}
\def\Eqs{Equations}
\def\beq{\begin{equation}}
\def\eeq{\end{equation}}
\defcitealias{Skoutnev_2024}{SB24}

\begin{document}

\title{Tayler Instability and Its Saturation in Rotating Stars}

\author{Valentin A. Skoutnev}
\affiliation{Physics Department and Columbia Astrophysics Laboratory, Columbia University, 538 West 120th Street New York, NY 10027,USA}

\author{Andrei M. Beloborodov}
\affiliation{Physics Department and Columbia Astrophysics Laboratory, Columbia University, 538 West 120th Street New York, NY 10027,USA}
\affiliation{Max Planck Institute for Astrophysics, Karl-Schwarzschild-Str. 1, D-85741, Garching, Germany}



\begin{abstract}

In stably stratified regions of rotating stars, magnetic instabilities play a key role by enabling dynamo and angular momentum transport. In particular, the Tayler instability of wound-up toroidal fields has long been considered to drive the Tayler-Spruit dynamo. However, the nonlinear saturation of the instability is poorly understood, despite its central role in the dynamo loop. In this paper, we use global simulations to verify the recently completed linear theory of the Tayler instability and determine its saturation. We confirm the picture of overstability of magnetostrophic and inertial waves, which grow on similar magnetostrophic timescales but oscillate with different frequencies. We find that magnetostrophic waves saturate with large amplitudes $\delta B/B\sim 1$ and dissipate the toroidal field on magnetostrophic timescales, while inertial waves saturate with $\delta B/B\ll 1$ and erode the toroidal field over much longer timescales. Moreover, magnetostrophic waves efficiently regenerate axisymmetric poloidal fields through nonlinear couplings while inertial waves do not. Our results indicate that the Tayler-Spruit dynamo can operate only through the magnetostrophic wave branch and that its steady-state may follow the scenario proposed by \cite{spruit2002dynamo}, ruling out the small-amplitude saturation scenario proposed by \cite{fuller2019slowing}. Implications for  angular momentum transport and magnetic field amplification in evolving stars and nascent neutron stars are discussed.

\end{abstract}



\section{Introduction}

The Tayler-Spruit dynamo (TSD)  is a candidate mechanism for angular momentum (AM) transport and magnetic field amplification in differentially rotating stars \citep{spruit2002dynamo}. It operates via the Tayler instability (TI) of strong toroidal fields \citep{tayler1973adiabatic}, which are generated by the winding of weak poloidal fields. The nonlinear saturation of the TI, accompanied by regeneration of the poloidal fields, closes the dynamo loop.

The TSD has been invoked to explain efficient AM transport inside evolving stars \citep{cantiello2014angular,aerts2019angular,fuller2019slowing}, predict the natal spins of compact objects \citep{fuller2019most,ma2019angular,gottlieb2024she}, and magnetize  nascent neutron stars \citep{margalit2022angular,barrere2022,reboul2025tayler,ng2026spinning}. However, the efficiency of the TSD remains highly uncertain due to the poorly understood properties of the TI, including its saturation and ability to regenerate poloidal fields.

The linear theory that determines the instability conditions and identifies the unstable modes was recently completed in \cite{Skoutnev_2024} \citepalias{Skoutnev_2024}. The analysis demonstrated that the TI in a rotating star is typically diffusively enabled: instability requires a dominant diffusive process to help overcome stabilization by the Coriolis force. As a result, the TI develops on length scales where diffusive and Coriolis timescales are comparable. This picture was not evident from the original marginal stability analysis in \cite{spruit1999differential}, and numerically testing the new linear theory is one of the goals of this paper.
 
Exponential growth of the unstable modes is followed by their saturation, which is strongly affected by rotation. When rotation is slow, TI saturation is well-understood: magnetic loops slip freely in opposite directions from the poles and saturate through parasitic Kelvin-Helmholtz instabilities on Alfv\'enic timescales $\sim 1/\omA$ \citep{ji2023magnetohydrodynamic}. This generates magnetic field fluctuations comparable in magnitude to the background field $\delta B\sim B$ that dissipate through a turbulent cascade \citep{bonanno2012breakdown,weber2015tayler,guerrero2019global,meduri2025gravity}. The TSD, however, operates in the limit of fast rotation ($\Omega \gg \omA$) where Coriolis forces restrain latitudinal slipping on Alfv\'enic timescales. In this case, the unstable eigenmodes are overstable waves---they oscillate with a real frequency while growing in amplitude on the magnetostrophic timescale $\sim \Omega/\omA^2$ \citep{pitts1985adiabatic,spruit1999differential,bonanno2013rotational}. Their saturation amplitude is a key unresolved question, as it controls the dynamo outcome and AM transport.

Two theoretical models predict different saturation levels based on disjoint possibilities. \cite{spruit2002dynamo} argued that Tayler modes saturate through large-scale, latitudinal displacements of magnetic loops in the unstable polar regions. In this picture, toroidal fields migrate across the pole  on magnetostrophic timescales and easily reconnect with opposite polarity fields, resulting in large saturation amplitudes $\delta B/ B\sim1$ and efficient dissipation of the background toroidal field. \cite{fuller2019slowing} instead argued that saturation occurs with small displacements of magnetic loops. Then, energy transfer can occur only through interactions between local perturbations of the background field, resulting in a lower saturation level $\delta B/ B\sim \omA/\Omega\ll1$ and less efficient dissipation. The lower saturation level and reduced dissipation predicted in \cite{fuller2019slowing} relative to \cite{spruit2002dynamo} leads to a TSD model that builds up stronger dynamo fields and transports AM more rapidly.

Distinguishing between these two possibilities has been a challenge for numerical models. Simulations of the rotating TI in local cylindrical geometry find alternative scalings for $\delta B/B$ \citep{ji2023magnetohydrodynamic}, while global simulations report results inconsistent with the predictions of linear theory \citep{monteiro2023global}. The simulations were, however, limited by either modest rotation rates \citep{ji2023magnetohydrodynamic} or uncontrolled, grid-based, diffusivities \citep{monteiro2023global}. Numerical effort has also gone into exploring the complete TSD loop by including differential rotation \citep{petitdemange2023spin,barrere2023numerical,barrere2026coexisting}, with a setup that brings in additional complexity, including turbulence excited by boundary layer instabilities and competing magnetic instabilities such as the magnetorotational instability.

In this paper, we aim to (1) numerically verify the linear instability theory developed in \citetalias{Skoutnev_2024} and (2) resolve the question of TI saturation. Our global numerical model of a stably stratified fluid ball is described in Section~\ref{sec:Methods}. Tests of the linear theory are presented in Section~\ref{sec:LinearTheory}, using simulations with different diffusivity ratios to isolate different TI modes. These modes are either overstable magnetostrophic waves (MW) or inertial waves (IW); the unstable IWs in particular are a new prediction of \citetalias{Skoutnev_2024}. We then examine their saturation and find qualitatively different outcomes. The unstable MW saturate with large-scale loop displacements, large amplitudes $(\delta B\sim B)$, and dissipate the background field on instability timescales, consistent with the picture in \cite{spruit2002dynamo}. By contrast, the unstable IW saturate with  smaller displacements, smaller amplitudes $(\delta B\ll B)$, and slower dissipation of the background field, similar to the picture in \cite{fuller2019slowing} originally proposed for MW. These findings are paralleled in a complementary paper, \cite{fraser2026}, that uses local simulations in a cylindrical geometry. In Section~\ref{sec:Theory}, we present analytical arguments that account for both MW and IW saturation by examining the nonlinear timescales for energy transfer between different modes. Implications for the TSD and AM transport in evolving stars and nascent neutron stars are discussed in Section~\ref{sec:Conc}.

\section{Methods}
\label{sec:Methods}

\subsection{Governing equations}

The magnetohydrodynamic (MHD) equations governing the typical subsonic fluid motions in stellar radiative zones can be simplified using the Boussinesq approximation \citep{spiegel1960boussinesq,spiegel1982magnetic}. In a frame rotating uniformly with rate $\boldsymbol{\Omega}=\Omega\hat z$, the MHD Boussinesq equations for the velocity $\bu$, magnetic field $\bB$, and temperature perturbation $T'$ from the background profile $T(r)$ are:
\begin{align}
\label{eq:MHD1}
    \frac{\partial \bu}{\partial t}+\bu\cdot \nabla \bu&=-\frac{\nabla P'}{\rho } 
    -\frac{\rho'}{\rho }
    g(r)\hat{r}-2\Omega\hat z\times\bu\nonumber\\
    &+\frac{(\nabla\times\bB)\times\bB}{4\pi\rho}+\nu\nabla^2\bu,\\
\label{eq:MHD2}
    \frac{\partial T'}{\partial t}+\bu\cdot \nabla T'&=-(d_rT-d_rT_{\rm ad})\bu\cdot\hat r+\kappa\nabla^2T',\\
\label{eq:MHD3}
    \frac{\partial \bB}{\partial t}&=\nabla\times(\bu\times \bB)+\eta\nabla^2\bB,\\
\label{eq:MHD4}
    \nabla \cdot \bB&=0,\\
\label{eq:MHD5}
    \nabla \cdot \bu&=0,
\end{align}
where $\nu$ is the  viscosity, $\eta$ is the magnetic diffusivity, $\kappa$ is the thermal diffusivity, $\rho $ is a constant reference density, $P'$ is the pressure perturbation, $g(r)$ is the gravitational acceleration, $d_r T_{\rm ad}$ is the adiabatic temperature gradient, and $\hat r$ is the radial unit vector in spherical coordinates $(r,\theta,\phi)$. The buoyancy force in the Boussinesq approximation depends only on the temperature perturbation: $\rho'g/\rho =-\alpha g T'/T$, where $\alpha=-(\partial \ln\rho/\partial \ln T)|_P$. The pressure $P'$ contributes to $\rho'$ at higher order and effectively acts as a constraint variable that enforces incompressibility of the fluid $\nabla \cdot \bu=0$.

\subsection{Background Magnetic field}

We are interested in the TI of an initially axisymmetric toroidal field $\bB_0=B_{\phi}(r,\theta)\hat \phi$ with a spatial profile representative of fields generated by differential rotation. For the approximately shellular differential rotation $\Omega(r)$ expected in stars \citep{zahn1992circulation}, winding $\partial_tB_\phi=B_r(r,\theta)\sin\theta r\partial_r\Omega(r)$ of the dipolar component of the radial field $B_r\propto \cos\theta$ imposes the large-scale angular dependence $B_{\phi}\propto \sin\theta\cos\theta$. While the accompanying radial dependence may generally be complicated, we adopt a simple linear profile:
\begin{align}
\label{eq:B_0}
    \boldsymbol{B}_0&=B_0\frac{r}{R}\sin\theta\cos\theta\,\hat\phi,
\end{align}
where $R$ is a characteristic radius of the radiative zone. The associated toroidal Alfv\'en frequency is defined as
\begin{align}
\label{eq:omA}
    \omA(r,\theta)&\equiv\frac{|B_{\phi}(r,\theta)|}{\sqrt{4\pi\rho }r\sin\theta}=\frac{B_0}{\sqrt{4\pi\rho }R}|\cos\theta|.
\end{align}
It is independent of radius and attains a maximum value $\omA=B_0/\sqrt{4\pi\rho }R$ on the polar axis ($\theta=0,\pi$). The Alfv\'en speed is then $v_{\rm A}=B_0/\sqrt{4\pi\rho}=\omA R$.

The toroidal field $\bB_0$ is assumed to be in magnetostatic equilibrium, satisfying the force balance between the pressure, buoyancy, and Lorentz forces:
\begin{align}
\label{eq:MagEq}
    0&=-\nabla P'_0+\alpha \frac{T'_0}{T}\rho g \hat{r}+\frac{(\nabla\times\bB_0)\times\bB_0}{4\pi}  .  
\end{align}
The irrotational and solenoidal parts of the Lorentz force are balanced by the pressure gradient and the buoyancy force, respectively. The equilibrium buoyancy force can be found after taking the curl of \Eq~(\ref{eq:MagEq}):
\begin{align}
\label{eq:curl_MagEq}
    \frac{1}{r}\frac{\partial }{\partial\theta}\left(\alpha\frac{T'_0}{T}\rho g\right)=\frac{\hat\phi\cdot\nabla\times((\nabla\times\bB_0)\times\bB_0)}{4\pi}.
\end{align}

\subsection{Numerical model}

We model a stellar radiative zone as a ball of radius $R$ and solve \Eqs~(\ref{eq:MHD1})-(\ref{eq:MHD5})  using the Dedalus pseudo-spectral code \citep{vasil2019tensor,lecoanet2019tensor,burns2020dedalus}. The initial equilibrium fields are determined from \Eqs~(\ref{eq:B_0}), (\ref{eq:MagEq}), (\ref{eq:curl_MagEq}), and $\bu_0=0$. The boundary conditions imposed at $r=R$ are stress-free for the velocity field, Dirichlet for the temperature $T'(r=R)=0$, and perfect conductor for the magnetic field.  The profiles of the gravitational acceleration and  \BV~frequency are chosen to have the forms 
\beq
  g(r)=g_0\,\frac{r}{R}, \qquad N^2(r)=2N_0^2\,\frac{r}{R},
\eeq
approximating the core of a stellar interior \citep{barker2010internal}. A convenient non-dimensionalization of the equations uses the ball radius for the length scale $\tilde r=r/R$ and Alfv\'enic units for the magnetic field $\tilde \bB=\bB/B_0$, velocity field $\tilde \bu=\bu/v_{\rm A}$, and timescale $\tilde t=\omA t$, with $\omA$ taken on the polar axis (\Eq~\ref{eq:omA}). The non-dimensional MHD Boussinesq equations are then:
\begin{align}
\label{eq:ndMHD1}
    \frac{\partial \tbu}{\partial\tilde t}+\tbu\cdot \tnabla \tbu&=-\tnabla \tilde P
    +\tvart\tilde r\hat{r}-\frac{2\Omega}{\omA}\hat z\times\tbu\nonumber\\
    &+(\tnabla\times\tbB)\times\tbB+\frac{Pm}{Lu}\tnabla^2\tbu,\\
\label{eq:ndMHD2}
    \frac{\partial \tvart}{\partial \tilde t}+\tbu\cdot \tnabla \tvart&=-\frac{N^2(\tilde r)}{\omA^2}\tbu\cdot\hat r+\frac{Pm}{PrLu}\tnabla^2\tvart,\\
\label{eq:ndMHD3}
    \frac{\partial \tbB}{\partial \tilde t}&=\tnabla\times(\tbu\times \tbB)+\frac{1}{Lu}\tnabla^2\tbB,\\
\label{eq:ndMHD4}
    \tnabla \cdot \tbB&=0,\\
\label{eq:ndMHD5}
    \tnabla \cdot \tbu&=0,
\end{align}
where $ \tnabla=R\nabla$, $\tilde P=P'/\rho v_{\rm A}^2$, and $\tvart= \alpha g_0RT'/(Tv_{\rm A}^2)$ is the buoyancy variable. The dimensionless control parameters are the ratios of the dynamical frequencies, the Lundquist number, and the thermal and magnetic Prandtl numbers: 
\beq
\frac{N_0}{2\Omega},\quad \frac{2\Omega}{\omA},\quad Lu=\frac{\omA R^2}{\eta},\quad Pr=\frac{\nu}{\kappa},\quad Pm=\frac{\nu}{\eta}.
\eeq

Simulations are initialized with additional Gaussian random noise for the velocity and magnetic fields with small amplitudes $|\tilde u|\sim|\tilde B|\ll1$ to seed the TI. The equations are integrated using a second order Runge-Kutta timestepper \citep{ascher1997implicit} and nonlinear terms are dealiased using the standard $3/2$ rule. The simulations have a resolution of  $n_{\max}$ for the highest order Jacobi polynomial used in the radial expansion and $\ell_{\max}=n_{\max}$ for the highest order spherical harmonic in the angular expansion. The most demanding simulations use $n_{\max}=384$.

\subsection{Diagnostics}

Our main diagnostics quantify the amplitudes of  perturbations, $\bu$ and $\bB-\bB_0$, relative to the initial equilibrium magnetic field, $\bB_0$. Since the TI eigenmodes are non-zero only at high latitudes (in particular, the WKB growth rate of the MW is positive for $\cos^2\theta>1/2$), it is appropriate to focus on the  polar regions $\theta<\theta_1$ and $\theta>\pi-\theta_1$. We will measure integrated quantities in these regions using a weight function $W(\theta)$ of the form
\begin{align}
    W(\theta)=1+\frac{1}{2}\left[\tanh\left(\frac{\theta-\pi+\theta_1}{\Delta\theta}\right)-\tanh\left(\frac{\theta-\theta_1}{\Delta\theta}\right) \right],
\end{align}
with $\theta_1=\pi/6$ and $\Delta\theta=\theta_1/5$.

The weighted, root-mean-square amplitudes of the perturbations are defined as:
\begin{align}
    \frac{\delta B(t)}{B_0}\equiv\left(\frac{\int |\bB-\bB_0|^2W(\theta)dV}{\int |\bB_0|^2W(\theta)dV}\right)^{1/2},\\
    \frac{\delta u(t)}{v_A}\equiv\left(\frac{\int 4\pi\rho |\bu|^2W(\theta)dV}{\int |\bB_0|^2W(\theta)dV}\right)^{1/2}.
\end{align}
These definitions analogously extend to individual components of the fields. The weighted amplitude of the axisymmetric field components is defined as
\begin{align}
    \frac{\overline B_j(t)}{B_0}\equiv\left(\frac{\int |\int B_j d\phi|^2W(\theta)dV}{(2\pi)^2\int |\bB_0|^2W(\theta)dV}\right)^{1/2},
\end{align}
where $j=r,\theta,\phi$.

\section{Confirming Linear Theory}

In this section, we use global simulations to test the analytical theory of the rotating TI developed in \citetalias{Skoutnev_2024}.

\label{sec:LinearTheory}
\subsection{Review of linear theory}

The TI of a toroidal magnetic field configuration reduces magnetic energy by rearranging the nested and stacked magnetic loops through slip and kink motions. However, in a rapidly rotating star with $2\Omega> \omA$, Coriolis forces restore motions perpendicular to the rotation axis. When plasma diffusivities $\kappa$, $\nu$, $\eta$ are equal (as can happen in simulations with diffusivities set by numerical dissipation), rotation stabilizes the TI, and any external perturbation excites waves that damp away. Instability is enabled when diffusivities differ: excited waves can become overstable, oscillating with frequency $\omega_r$ while their amplitude grows with rate $\gamma>0$. The unstable wave modes are categorized as either MW or IW based on their frequency (see below). The most unstable eigenmodes have global latitudinal and azimuthal structure, but have short radial scales where diffusive processes operate sufficiently fast to compete with the Coriolis force.  

Unstable eigenmodes in the WKB approximation have the form $\propto\exp[i(k_r r + \ell \theta + m\phi-\omega t)]$ with complex frequency $\omega=\omega_r+i\gamma$, large radial wavenumbers $k_r\gg k_\theta\sim \ell/r$, low angular modes $\ell\sim{\cal O}(1)$, and $|m|=1$ as the only unstable azimuthal mode. Each diffusive process across the short radial scale $k_r^{-1}\approx k^{-1}$ has a timescale:
\beq
    t_{\kappa}=\frac{\kappa k^4}{k_\theta^2 N^2},\quad  t_\nu=\frac{1}{\nu k^2},\quad t_\eta=\frac{1}{\eta k^2}.
\eeq
The TI growth rate peaks at wavenumbers $k=\kTI$ where the rotation timescale  $t_\Omega=(2\Omega)^{-1}$ is comparable to a diffusion timescale. Equating $t_\Omega$ with $t_{\kappa}$, $t_\nu$, $t_\eta$ gives three canonical wavenumbers 
\begin{align}
\label{eq:kTI}
    & k_{\kappa}=\left(\frac{k_\theta^2N^2}{\om\kappa }\right)^{1/4},\quad k_\nu=\left(\frac{\om}{\nu}\right)^{1/2}, \quad k_\eta=\left(\frac{\om}{\eta}\right)^{1/2}.
\end{align}
The maximum possible growth rate at $\kTI=k_\nu,k_\eta$ is $\gamma = \omA^2/4\Omega$. The instability at $\kTI=k_\kappa$ is slightly slower, with $\gamma \approx \omA^2/5\Omega$. 

The properties of the unstable eigenmodes are determined by their wave type. The eigenmodes categorized as MW have low frequencies $|\omega_r|\sim \omA^2/4\Omega\ll\omA$ and are magnetically dominated: $\delta B/\sqrt{4\pi \rho}\, \delta u\sim 2\Omega/\omA\gg 1$. The $\kkappa$ and $\knu$ modes of TI are overstable MW. Conversely, IW have high frequencies $|\omega_r|\sim \om\gg\omA$ and are kinetically dominated: $\delta B/\sqrt{4\pi \rho} \,\delta u\sim \omA/2\Omega$. The $\keta$ mode of TI is an overstable IW.

The instability conditions for each $\kTI\in\{k_\kappa,k_\nu,k_\eta\}$ are controlled by diffusivities. Instability at $\kTI=k_\kappa$ requires both negligible magnetic diffusion at $k_\kappa$ ($\eta k_\kappa^2\ll\gamma $) and fast thermal diffusion at the stratification wavenumber $k_N=k_\theta N/\omA$ ($\kappa k_N^2\gg\gamma $). Combining these requirements gives the condition
\beq
\label{eq:kkappa_cond}
    \left(\frac{N}{\om} \frac{\eta}{\kappa}\right)^{1/2}\left(\frac{\kappa k_\theta^2}{\om}\right)^{1/4}\ll\frac{\omA}{\om}\ll\left(\frac{N}{\om} \right)^{1/2} \left(\frac{\kappa k_\theta^2}{\om}\right)^{1/4}.
\eeq
Instability at $\kTI=k_\nu$ only requires negligible magnetic diffusion $\eta k_\nu^2\ll\gamma$, yielding the condition
\beq
\label{eq:knu_cond}
    \frac{\omA}{2\Omega}\gg 
    \frac{1}{Pm^{1/2}}.
\eeq
Similarly, instability at $\kTI=k_\eta$ only requires negligible viscosity $\nu k_\eta^2\ll\gamma $, yielding the condition
\beq
\label{eq:keta_cond}
    \frac{\omA}{2\Omega}\gg Pm^{1/2}.
\eeq
The above conditions imply that instability of the MW modes $\kkappa$ and $\knu$ is possible when $\kappa,\nu\gg \eta$, while instability of the IW mode $\keta$ is possible when $\eta\gg\nu$. Therefore, the two types of TI modes, MW and IW, never occur simultaneously.

We note that the linear theory for the MW branch changes for toroidal fields with steeper gradients near the axis compared to our setup with $B_\phi\propto r\sin\theta$. In particular, the change occurs for $B_\phi\propto (r\sin\theta)^p$ with $p>3/2$. This can more easily occur when differential rotation is along cylinders, such as inside stars rotating near break up or merger remnants. For such scenarios, the MW branch is no longer diffusively enabled. It is unstable with a constant magnetostrophic growth rate across a range of wavenumbers $k_1\leq k\leq k_2$ (see \citetalias{Skoutnev_2024}), instead of at $\kkappa$ and $\knu$ individually. We expect that TI saturation in this $p>3/2$ case is similar to the $p<3/2$ case since dynamics at all unstable wavenumbers remains magnetostrophic.

\begin{figure*}
    \centering
    \includegraphics[width=\linewidth]{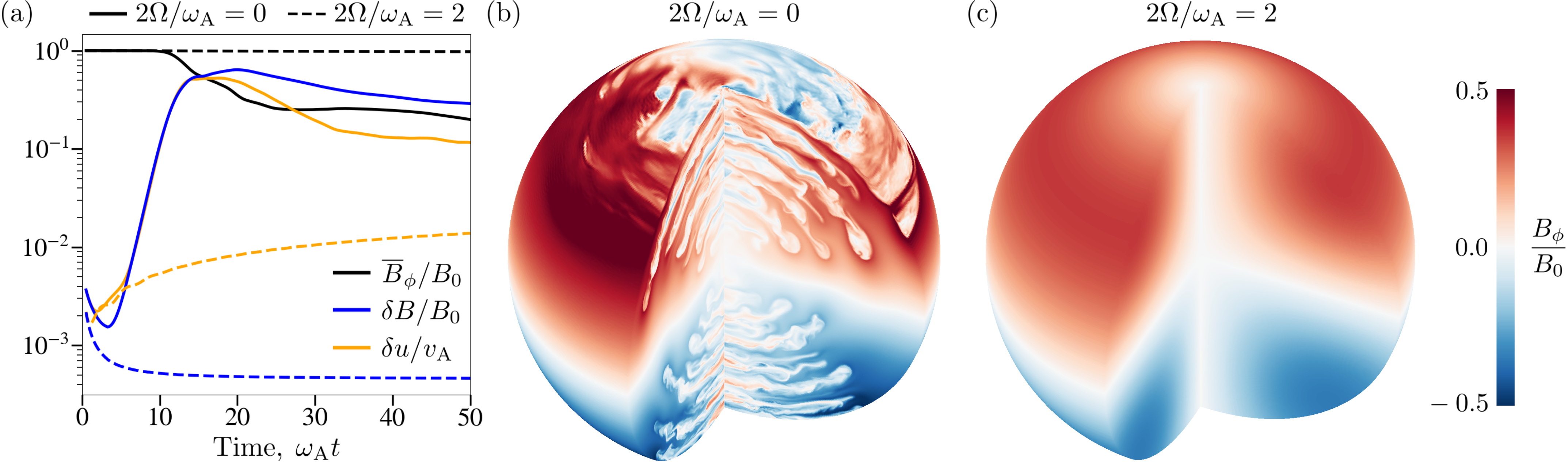}
    \caption{Comparison of non-rotating and rotating simulations with equal diffusivities $(\nu=\eta=\kappa)$ to demonstrate the stabilizing effect of rotation predicted by linear theory. Panel (a) tracks the evolution of the background and perturbation field amplitudes, comparing the non-rotating (solid) and rotating (dashed) simulations. Panels (b) and (c) show the toroidal field component $B_\phi$ at time $\omA t=16$ when the TI is saturating in the non-rotating simulation. The visualized ball shows the radial interval $0\leq r\leq 0.8R$ and meridional slices at $\phi=0,3\pi/2$. }
    \label{fig:NR}
\end{figure*}

\begin{figure*}
    \centering
    \includegraphics[width=\linewidth]{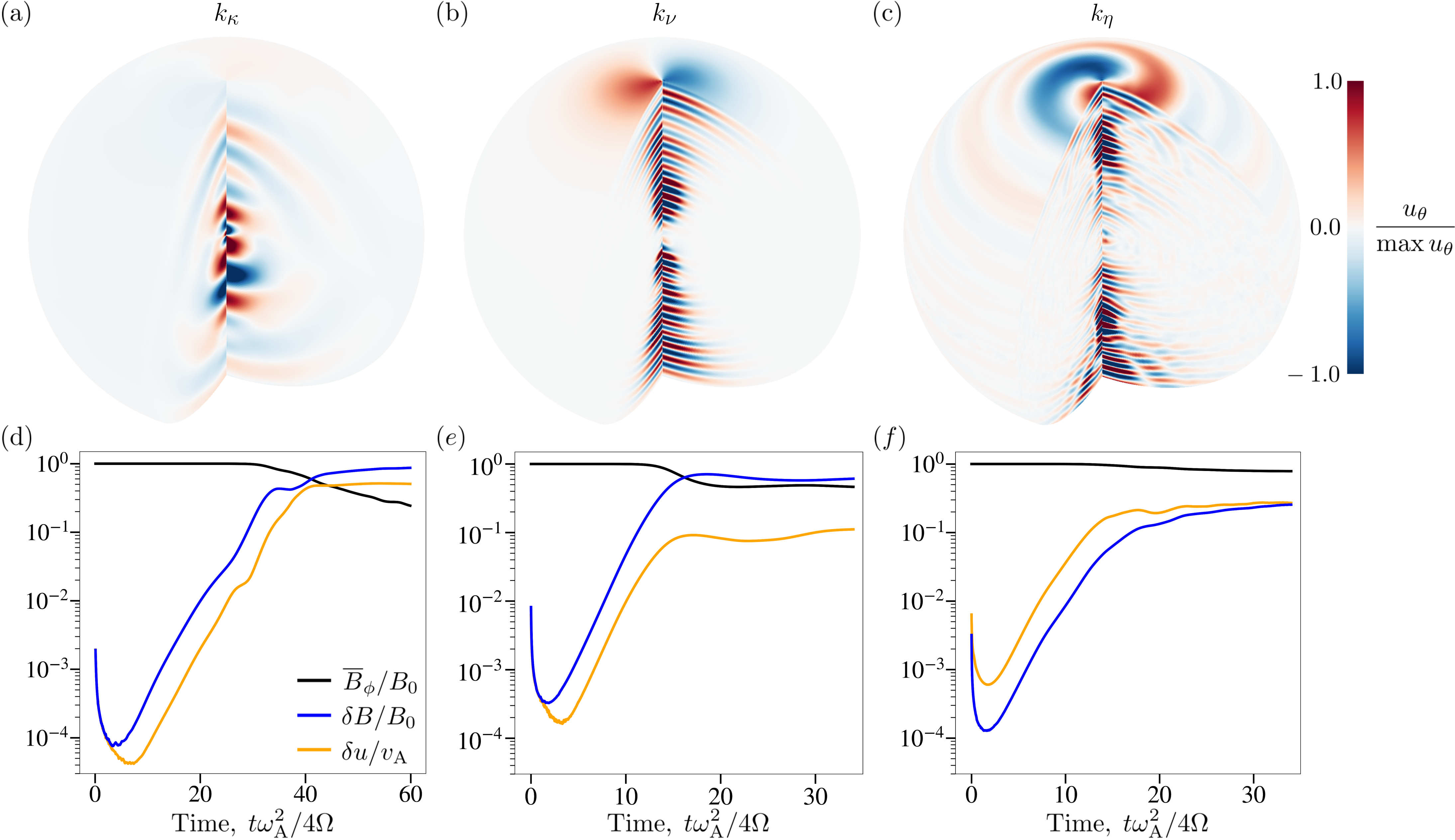}
    \caption{
    Validation of linear theory using rotating simulations that selectively destabilize the $k_\kappa$, $k_\nu$, and $k_\eta$ modes by varying diffusivity ratios (see Table~\ref{tab:FiducialSetofThree}). The results of the three simulations are shown in three columns labeled as $k_\kappa$, $k_\nu$, and $k_\eta$. Top row visualizes the velocity component $u_\theta$ during the linear phase of the TI when perturbations amplitudes are small, $|u_\theta|/v_{\rm A}={\cal O}(10^{-3})$, and the TI eigenmodes are growing exponentially. Bottom row tracks the evolution of the field amplitudes. MWs are overstable in the $k_\kappa$ and $k_\nu$ simulations and IWs are overstable in the $k_\eta$ simulation. }
    \label{fig:Linear}
\end{figure*}

\subsection{Effect of rotation on stability}

We first verify the stabilizing effect of rotation by comparing simulations with $\Omega=0$ and $2\Omega/\omA=2$. Both simulations have equal diffusivities $\nu=\eta=\kappa$, stable stratification with $N_0/\omA=4$, and Lundquist number $Lu=10^5$. Linear theory predicts instability in the non-rotating case, provided that $\omA>\omega_{\rm A,c}=(4\eta k_\theta^2N^2/3)^{1/3}$ with $k_\theta\sim 5/R$ (\cite{spruit1999differential}; \citetalias{Skoutnev_2024}).\footnote{Tayler modes grow at the poles ($|\cos\theta|>1/\sqrt{2}$ for MW and $|\cos\theta|>\sqrt{3}-1$ for IW), corresponding to angular harmonics $\ell\gtrsim4$ and $k_\theta\sim \sqrt{\ell(\ell+1)}/R\sim5/R$.} Our simulation has $\omA\approx 6\omega_{\rm A,c}$ and should give a violent instability. In the rotating case, no TI is expected because $\nu=\kappa=\eta$. These expectations are confirmed by the simulation results presented in Figure~\ref{fig:NR}.

The non-rotating global simulation agrees well with local simulations in a cylindrical geometry \citep{ji2023magnetohydrodynamic}. The TI grows on Alfv\'enic timescales $t\sim\omA^{-1}$ as field loops slide away from the poles on radial scales smaller than the stratification scale $2\pi/k_N\approx R/4$.\footnote{Global modes of the non-rotating TI have recently been found to grow faster than the polar, WKB modes when the system is near marginal stability $\omA\approx \omega_{\rm A,c}$ \citep{meduri2025gravity,gusakov2026beyond}. WKB modes dominate in our simulations because the chosen $\omA$ significantly exceeds $\omega_{\rm A,c}$.} Saturation begins once shear between radially-adjacent loops excites the Kelvin-Helmholtz instability. This occurs when the Richardson number $Ri\sim N^2/(k_N\delta u)^2$ reaches $\sim 1$,
resulting in Alfv\'enic saturation $\delta u\sim \delta B/\sqrt{4\pi \rho }\sim \overline B/\sqrt{4\pi \rho }$  and turbulent dissipation on Alfv\'enic timescales.

By contrast, in the rotating simulation with the same $\nu=\eta=\kappa$, seed perturbations do not trigger instability; instead, the seed MW and IW decay away. The long-term evolution is governed by gradual thermal and magnetic diffusion of the background buoyancy and magnetic fields, which induces slow meridional flows with $\delta u\ll \overline B/\sqrt{4\pi\rho }$ (not shown in Figure~\ref{fig:NR}). 

\subsection{Linear phase of the rotating TI}

\begin{table}[]
    \centering
    \begin{tabular}{c|c|c|c|c|c}
        Simulation & $N_0/2\Omega$ & $2\Omega/\omA$  & $Pm$ & $Pr$ & $Lu$ \\
        \hline
         $k_\kappa$ & 8 & 3 & $1$ & $400^{-1}$ & $3\cdot10^4$\\
         $k_\nu$ & 2 & 3 &$225$ & $225$ & $7\cdot10^5$\\
         $k_\eta$ & 2 & 3 &$225^{-1}$ & $1$ & $3\cdot10^3$
    \end{tabular}
    \caption{Parameters of the three simulations presented in Figures~\ref{fig:Linear} and~\ref{fig:NL}. Simulations maintain the same ordering of dynamical frequencies $\omA<2\Omega<N$ as stellar interiors.}
    \label{tab:FiducialSetofThree}
\end{table}

\begin{figure*}
    \centering
    \includegraphics[width=\linewidth]{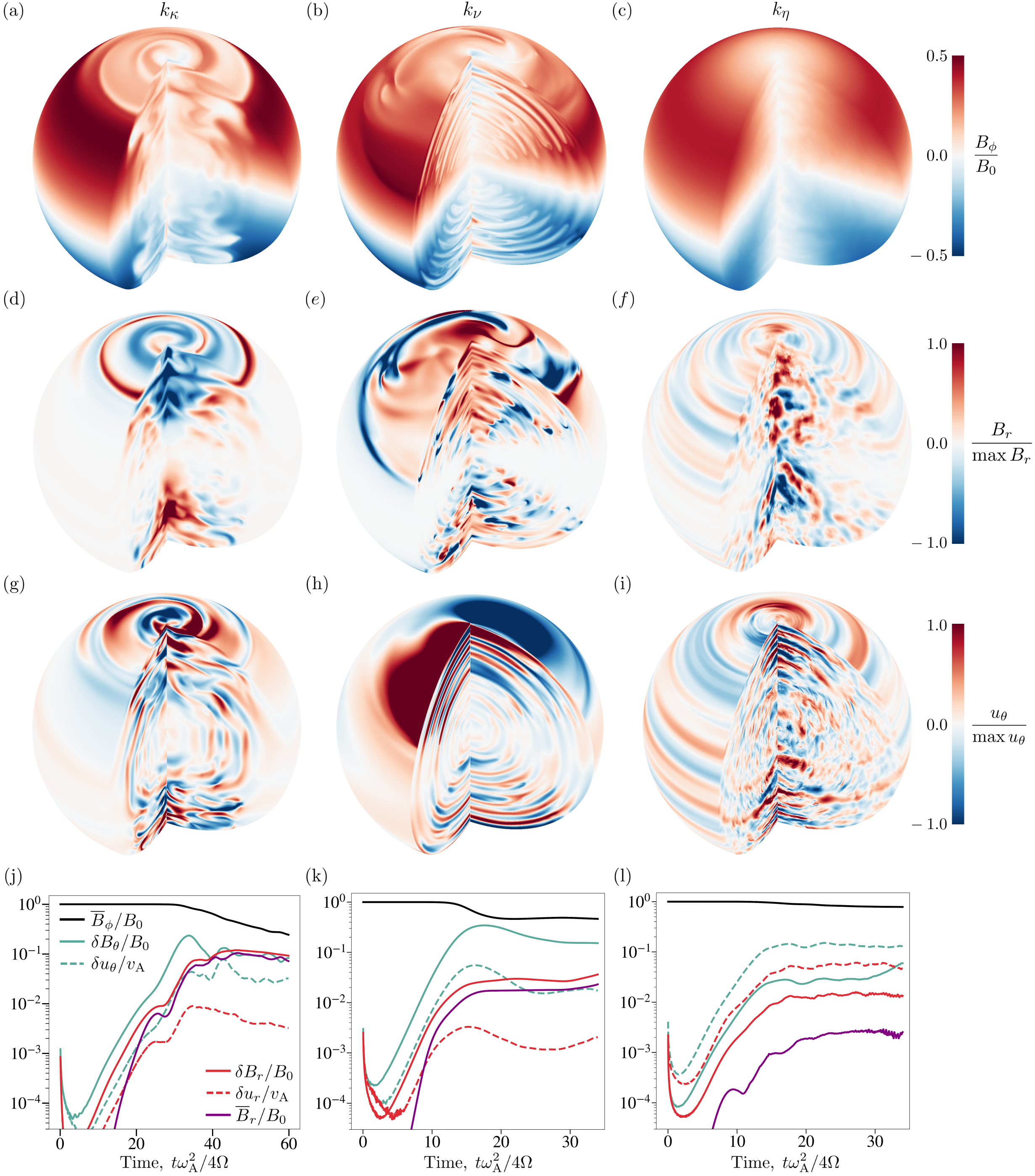}
    \caption{
    TI saturation in the $k_\kappa, k_\nu,$ and $k_\eta$ simulations following the linear phase shown in Figure~\ref{fig:Linear}. Simulations are organized by column and visualized at times $t\omA^2/4\Omega=40$, $20$, and $20$, respectively. Top three rows correspond to different components of the fields normalized to their maximum values. Bottom row tracks the time evolution. MW saturation results in large-scale displacements of toroidal field loops and perturbations with $\delta B\sim B_0$, which destroys the  initial axisymmetry of the toroidal field as seen in panels (a) and (b). IW saturation is mild by comparison, with $\delta B\ll B_0$ and minor distortion of the initial toroidal field as seen in panel (c). 
    }
    \label{fig:NL}
\end{figure*}

The reliance of the rotating TI on diffusive processes becomes evident when the diffusivity ratios are varied. We present three fiducial rotating simulations with $2\Omega/\omA=3$ and diffusivities ordered as  (1) $\kappa\gg \nu=\eta$, (2) $\nu\gg \kappa=\eta$, and (3) $\eta\gg \kappa=\nu$, isolating the $\kTI=k_\kappa$, $k_\nu$, and $k_\eta$ modes of the TI, respectively. In these simulations, we remove the diffusion of the initial magnetic and buoyancy fields, $\bB_0$ and $\vartheta_0$, to avoid Ohmic or thermal decay during the slow growth of the rotating TI.\footnote{Any decrease of the background field energy during TI saturation is then cleanly attributed to the TI and not diffusion. This is implemented by solving the MHD Boussinesq equations with the substitution $\tilde \bB'(t)=\tilde \bB(t)-\tilde \bB_0$, $\tilde \vartheta'(t)=\tilde \vartheta(t)-\tilde \vartheta_0$, and dropping the diffusion operators that act on $\tilde \bB_0$ and $\tilde \vartheta_0$.} The simulation parameters are detailed in Table~\ref{tab:FiducialSetofThree} and the results are shown in Figures~\ref{fig:Linear} and \ref{fig:NL}. 

All three simulations develop exponentially growing eigenmodes during the linear phase of instability (Figures~\ref{fig:Linear}). They grow on the expected magnetostrophic timescales $\sim 4\Omega/\omA^2$, with the $k_\kappa$ modes growing somewhat slower than the $k_\eta$ and $k_\nu$ modes. The eigenmodes are spatially concentrated near the poles with a clear $|m|=1$ azimuthal structure characteristic of the TI. The observed number of radial wavelengths ($\sim 3$, $\sim 10$, and $\sim 11$  in panels (a), (b), and (c), respectively) agrees with the WKB estimates for the radial wavenumbers in \Eq~(\ref{eq:kTI}) ($0.8Rk_\kappa/2\pi\approx3$, $0.8Rk_\nu /2\pi=12$, and $0.8Rk_\eta /2\pi\approx12$).

The amplitude diagnostics for $\delta u$ and $\delta B$ (Figure~\ref{fig:Linear}) confirm the asymmetric partition of magnetic and kinetic energy in MW and IW eigenmodes: the MW are magnetically dominated (with measured $\delta B/\delta u\sqrt{4\pi\rho }\approx 4.5)$, while the IW are not (with $\delta B/\delta u\sqrt{4\pi\rho }\approx 1/4.5$).\footnote{The relation between the velocity and magnetic field, $\bu$ and $\bb$, of a TI eigenmode can be found from the linearized induction equation $\bb=\Re[-m\omA \bu\sqrt{4\pi\rho}/\omega_\eta]$. For MW modes where $\omega\approx\omA^2/4\Omega(1+i)$ and $\omA^2/4\Omega\gg\eta \kTI^2$, one expects $|\bb|/|\bu|\sqrt{4\pi\rho}\approx 2\sqrt{2}\Omega/\omA$. This equals $3\sqrt{2}\approx 4$ for our simulation parameters and agrees well with the measured value. A similar estimate can be made for the IW to find $|\bb|/|\bu|\sqrt{4\pi\rho}\approx \sqrt{2}\omA/4\Omega$.} The horizontal field components (Figure~\ref{fig:NL}) have the largest amplitudes ($\delta B_\phi$ and $\delta B_\theta$ for MW, and $\delta u_\phi$ and $\delta u_\theta$ for IW). The magnetic field components follow the scaling $\delta B_\phi\sim  (k_\theta R)\,\delta B_\theta \sim (\kTI R)\,\delta B_r$, which comes from $\nabla\cdot\bB=0$ with $k_\phi \delta B_\phi\sim k_\theta \delta B_\theta \sim \kTI \delta B_r$ and $k_\phi= 1/R$. Similar scalings apply to the velocity components due to the incompressibility constraint. Several of these eigenmode properties are imprinted onto the turbulence that develops in the subsequent saturation phase.

We also note that weak nonlinear couplings pump the axisymmetric radial field $\overline B_r$ to remarkably large amplitudes during the linear phase of the MWs, but not of the IW (Figure~\ref{fig:NL}). For both the $\kkappa$ and $\knu$ modes of the MW branch, the axisymmetric radial field grows to a significant fraction of the total radial field, $\overline B_r\lesssim \delta B_r$, by the end of the linear phase. This is consistent with observed growth of axisymmetric vertical fields $\overline B_z$ in cylindrical simulations \citep{ji2023magnetohydrodynamic}. For the $\keta$ mode of the IW branch, $\overline B_r\ll \delta B_r$ remains satisfied.

\begin{figure}
    \centering
    \includegraphics[width=\linewidth]{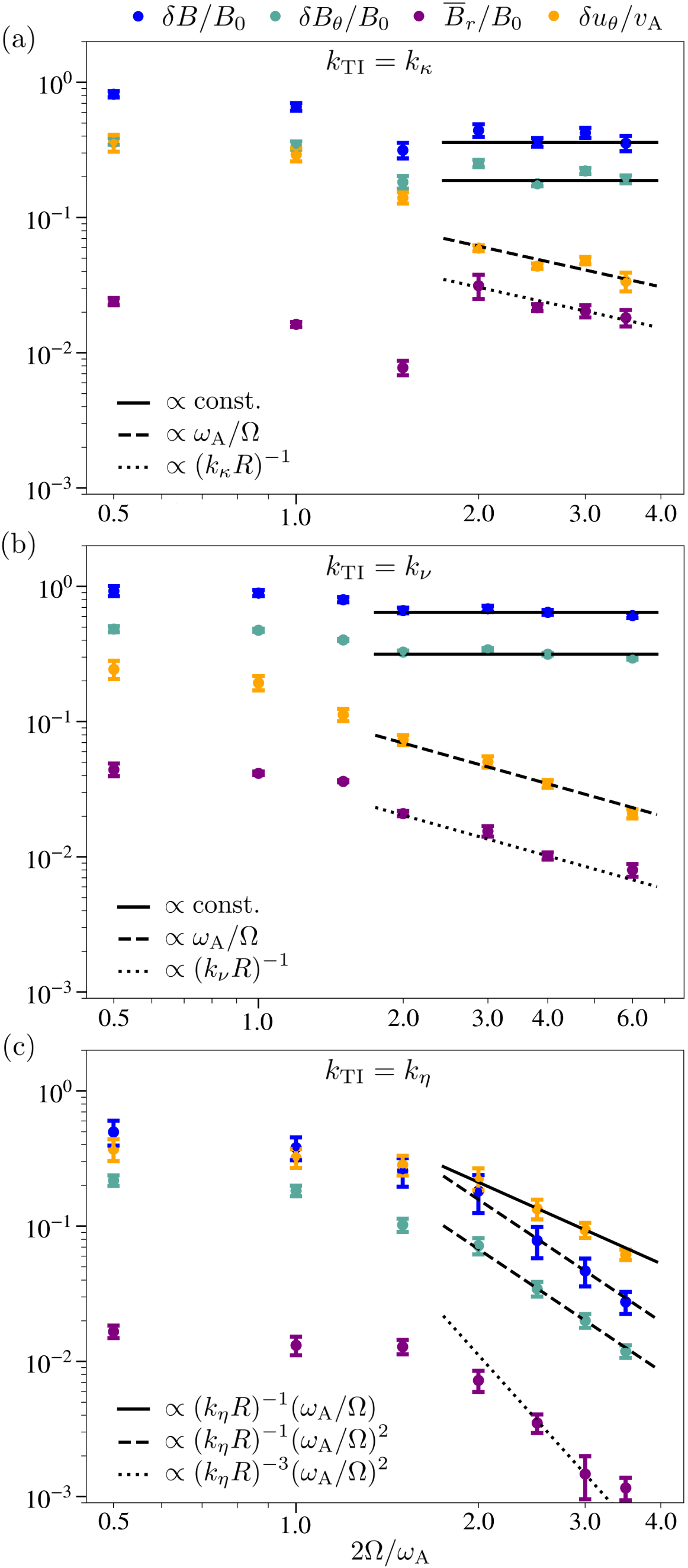}
    \caption{Amplitudes of the magnetic and velocity perturbations at TI saturation versus rotation parameter $2\Omega/\omA$. Results for the $\kkappa$, $\knu$, and $\keta$ modes of TI are presented in panels (a), (b), and (c), respectively. The colored points show the measured saturated amplitudes of $\delta B$, $\delta B_\theta$, $\delta u_\theta$, and $\overline{B}_r$, with normalizations indicated at the top of the figure. Their theoretical scalings developed in Section~\ref{sec:Theory} are indicated by the black lines. One can see that the MW instability (panels a,b) and the IW instability
    (panel c) give different scalings of the saturation amplitudes with rotation.
    }
    \label{fig:Scaling}
\end{figure}

\section{Nonlinear saturation of the rotating TI}
\label{sec:Sims_NL}

Saturation stops the exponential growth of the unstable eigenmodes. This occurs around $t\omA^2/4\Omega \gtrsim35$ in the $k_\kappa$ simulation, and $t\omA^2/4\Omega \gtrsim15$ in the $k_\nu$ and $k_\eta$ simulations. As seen in Figure~\ref{fig:NL}, the MW and IW have different saturation dynamics, which we discuss separately below.

The simulations demonstrate that MW instabilities (at $k_\kappa$ or $k_\nu$) saturate with magnetic perturbations comparable to the background field $\delta B\sim B_0$, and correspondingly large horizontal displacements\footnote{The magnitude of horizontal displacements $\xi_h\sim|\xi_\phi|\sim|\xi_\theta|$ follows from the horizontal components of the induction equation, e.g., $\partial_t \delta B_\phi\sim \bB_0\cdot \nabla \delta u_\phi$ with $\partial_t \xi_\phi\sim \delta u_\phi$ and $\bB_0\cdot \nabla \sim B_0k_\phi$.}  $\xi_h\sim R (\delta B/B_0)\sim R$. Over a few magnetostrophic timescales, magnetic loops in the unstable polar regions become displaced from the polar axis, breaking the axisymmetry of the initial field. Warping of the initially circular magnetic field lines and reconnection with opposite polarity field lines then drive a forward energy cascade, creating progressively smaller-scale structures that eventually dissipate energy at diffusive scales. 

The MW instabilities at $k_\nu$ and $k_\kappa$ exhibit different turbulent structures due to their disparate diffusivity ratios. In the $k_\nu$ case, the velocity field remains smooth on instability scales while the magnetic field develops small-scale reversals characteristic of high $Pm$ MHD turbulence (e.g. see \citealt{rincon2019dynamo}). In the $\kkappa$ case, magnetic and velocity fields have similar length scales since both the magnetic diffusivity and viscosity are small. In both cases, the axisymmetric radial field $\overline B_r$ generated during the linear phase is sustained or further amplified. 

Simulations of IW instability at $k_\eta$ show saturation with small amplitudes $\delta B\ll \delta u/\sqrt{4\pi \rho}\ll B_0$ and correspondingly small displacements $\xi_h\sim R (\delta B/B_0)\ll R$. The generated turbulence is characterized by structures in the velocity field much smaller than the instability scale, which we interpret in Section~\ref{sec:Theory} as secondary IW modes created by parametric instabilities of the IW amplified by the TI. The saturation with $\delta B\ll B_0$ leads to a quasi-steady state where the TI remains balanced by the nonlinear processes while the background field $B_0$ is gradually depleted. The depletion occurs on timescales much longer than the magnetostrophic timescale, which our simulations do not cover. We also observe that generation of axisymmetric radial field $\overline B_r$ during and after IW saturation remains inefficient compared to the $\overline B_r$ found during MW saturation. An analytical explanation of these results is given below.

The dependence of saturation amplitudes on the rotation rate is a key debated ingredient of TSD models. We have measured this dependence by running simulations with different values of the rotation parameter $2\Omega/\omA$. The measurements are done for each of the three possible modes of TI: $k_\kappa$, $k_\nu$, and $k_\eta$. The scan is performed at fixed super-criticality, maintaining a constant ratio between $\omA$ and its critical values for instability. Details of the scan are given in Appendix~\ref{ap:ParameterScan}.  Note that all the TI wavenumbers increase linearly with the rotation parameter in our scan: $\kkappa R,\knu R,\keta R\propto \Omega/\omA$.

The results are presented in Figure~\ref{fig:Scaling} and compared  with the analytical scalings developed in Section~\ref{sec:Theory}. The rotating regime is clearly distinguished for $2\Omega/\omA>1$ where the measured amplitudes are in good agreement with predicted scalings for both MW and IW. For MW modes, the magnetic amplitudes saturate with $\delta B/B_0\sim \delta B_\theta/B_0 \propto (2\Omega/\omA)^0$ and $\overline B_r/B_0\propto (\kTI R)^{-1}$ for both $\kkappa$ and $\knu$ simulations. The velocity amplitudes follow $\delta u/v_{\rm A}\propto(2\Omega/\omA)^{-1}$ closely in the $\knu$ simulations, while the $\kkappa$ simulations have greater scatter.\footnote{The $\kkappa$ simulations can develop $m=0$ azimuthal (``zonal") flows that are not affected by Coriolis forces and may not follow magnetostrophic scalings. Because of this, we use $\delta u_\theta$ as a proxy for the typical velocity fluctuations in Figure~\ref{fig:Scaling}.} For TI in the IW mode, the saturated velocity amplitudes are consistent with the scaling $\delta u/v_{\rm A}\propto (\keta R)^{-1}(2\Omega/\omA)^{-1}$, which is expected from parametric instabilities (Section~\ref{sec:Theory}). The saturated amplitude of magnetic field perturbations then follows the scaling $\delta B/B_0\propto (k_\eta R)^{-1}(2\Omega/\omA)^{-2}$. The expected theoretical scaling for the axisymmetric radial field, $\overline B_r/B_0\propto (k_\eta R)^{-3}(2\Omega/\omA)^{-2}$ appears to be slightly steeper than the data.

\section{Theory of saturation}
\label{sec:Theory}

The TI in a rotating star drives the growth of oscillatory eigenmodes (MW or IW) in the polar regions.  Saturation of the instability can be understood by decomposing the evolving fields based on their azimuthal mode numbers: $\bB= \bB_0+ \bB_1+ \bB_2+...$ and $\bu=\bu_0+ \bu_1+\bu_2+...$, where the subscript in $ \boldsymbol{B}_m$ indicates the component with azimuthal numbers $\pm m$. During the early, linear stage of the instability, the fields primarily consist of the background field with $m=0$ and the exponentially growing eigenmodes with $|m|=1$: $\bB\approx \bB_0+\bB_1$ and $\bu\approx \bu_1$ satisfying $\bB_1,\bu_1\sqrt{4\pi\rho}\ll|\bB_0|\approx B_0$. The eigenmodes, however, are not exact nonlinear solutions. Their nonlinear interactions can both extract energy from the background field and transfer energy to $|m|=2$ modes, which then can couple to even higher azimuthal modes. Additionally, the eigenmodes are standing waves, which are susceptible to secondary parametric instabilities that directly pump energy to higher wavenumbers (smaller scales). All three nonlinear processes accelerate as the amplitudes of unstable waves ($\bB_1$ and $\bu_1$) grow. Saturation begins once the timescale for at least one of these processes becomes comparable to the linear instability timescale. Below, we estimate the nonlinear timescales and  obtain predictions for the wave saturation amplitudes.

\subsection{Timescales}

Analytical estimates for nonlinear timescales can be made by examining the induction equation in the weakly nonlinear regime. Retaining the lowest order nonlinear terms, we write the induction equation for the first three azimuthal numbers $|m|=0,1,2$ as follows
\begin{align}
\label{eq:B0_evol}
    \partial_t \bB_0&+{\cal L}_0=\nabla\times( \langle\bu_1\times \bB_1\rangle_0),\\
    \label{eq:B1_evol}
    \partial_t \bB_1&+{\cal L}_1=\nabla\times( \langle\bu_1\times \bB_2\rangle_1+ \langle\bu_2\times\bB_1\rangle_1),\\
    \label{eq:B2_evol}
\partial_t \bB_2&+{\cal L}_2=\nabla\times( \bu_1\times \bB_1- \langle\bu_1\times \bB_1\rangle_0),
\end{align}
where $\langle...\rangle_m$ denotes projection onto modes with azimuthal number $|m|$, and ${\cal L}_m=-\nabla\times( \bu_m\times \bB_0+\bu_0\times \bB_m)-\eta\nabla^2 \bB_m$.
Linear theory considers the limit of infinitesimal perturbations, in which the nonlinear terms on the right hand side are neglected. The TI eigenmodes and eigenvalues are then determined from the resulting uncoupled equations, $\partial_t\bB_m+{\cal L}_m=0$ \citepalias{Skoutnev_2024}.

The nonlinear terms couple the evolution of  azimuthal modes. Induction by the eigenmodes $\propto \bu_1\times\bB_1$ can affect both $m=0$ and  $|m|=2$ modes (\Eqs~\ref{eq:B0_evol} and \ref{eq:B2_evol}). Interaction of the $|m|=2$ and $|m|=1$ modes then feeds back onto the $|m|=1$ modes (\Eq~\ref{eq:B1_evol}).  The rate of each process is determined by the magnitude of the corresponding nonlinear term. These equations are also coupled to an analogous azimuthal decomposition of the momentum and buoyancy equations, whose nonlinear Lorentz and advection terms can likewise couple the $|m|=1$ eigenmodes with the $m=0$ and  $|m|=2$ modes of the velocity field. Below, we obtain analytical estimates for the nonlinear timescales by considering the induction equation alone.

Let us first consider the evolution of the background ($m=0$) toroidal field in \Eq~(\ref{eq:B0_evol}). It will significantly change over a nonlinear timescale given by
\begin{align}
    \label{eq:t_nl_0}
    t_{0}^{-1}\sim\frac{|\hat \phi\cdot\nabla\times(\langle\bu_1\times\bB_1\rangle_0)|}{| B_{\phi,0}|}\sim |k_\phi u_{\phi,1}|\frac{ | B_{\phi,1}|}{B_0},
\end{align}
where we used incompressibility scalings $|k_r u_{r,1}|\sim |k_\phi u_{\phi,1}|$. The prefactor in \Eq~(\ref{eq:t_nl_0}) is $\sim{\cal O}(1)$, as we verify in Appendix~\ref{ap:NL}  by explicitly calculating the induction term using eigenmodes for $\bu_1$ and $\bB_1$. This nonlinearity weakens the background toroidal field that provides the energy reservoir for the TI, thereby suppressing the linear source term ${\cal L}_1$ driving the $|m|=1$ Tayler modes. 

We now turn to the sourcing of $|m|=2$ modes and their feedback onto the $|m|=1$ modes. Using $|\bu_1\times\bB_1|\propto e^{2\gamma t}$, a bound on the amplitude of the $|m|=2$ modes can be obtained from \Eq~(\ref{eq:B2_evol}) as
\begin{align}
    \label{eq:B_2_mag}
    |\bB_2|    \sim|\! \int dt\,\nabla\times(\bu_1\times\bB_1)|\lesssim\frac{|k_\phi u_{\phi,1}|}{2\gamma}|B_{\phi,1}|.
\end{align}
A similar estimate can be obtained from $|\bB_2| \sim|\! \int dt\,\nabla\times(\bu_2\times\bB_0)|$, where $\bu_2$ is sourced from the nonlinear Lorentz ($(\nabla\times\bB_1)\times  \bB_1/4\pi\rho$) or advection ($\bu_1\cdot \nabla  \bu_1$) terms in the momentum equation (Appendix~\ref{ap:NL}). The timescale for draining energy from $|m|=1$ modes via excitation of $|m|=2$ modes may be estimated from \Eq~(\ref{eq:B1_evol}) as follows
 \begin{align}
     \label{eq:t_nl_1}
    t_{\mathrm{cas}}^{-1}\sim\frac{|\nabla\times(\langle\bu_1\times\bB_2\rangle_1|}{| \bB_1|} \lesssim \frac{|k_\phi  u_{\phi,1}|^2}{\gamma} .
\end{align}
Higher chains of nonlinear interactions also source $|m|>2$ modes with growth rate $|m|\gamma$ and smaller amplitudes \citep{ji2023magnetohydrodynamic}. Physically, these couplings are the manifestation of the forward cascade transferring energy to smaller scales.

Finally, we estimate the timescale associated with parametric instabilities. The parametric instabilities of IW have been extensively studied for both standing \citep{Lifshitz1996,Miyazaki1998,Kerswell1999} and propagating waves \citep{Mora2021,AM2024,skoutnev2026stability}. IW are susceptible  to a form of parametric instability known as the triadic resonance instability, in which energy is transferred from the primary wave to a pair of daughter waves that satisfy certain resonance conditions. Because the daughter waves can have much larger wavenumbers, this process drives a non-local forward cascade drawing energy from the IW eigenmodes. The instability growth rate scales with the wave's total velocity amplitude and wavenumber. Its growth timescale $t_{\rm pi}$ is given by
\begin{align}
    t_{\rm pi}^{-1}\sim |\bu_1|k\sim |k_\phi u_{\phi,1}|\keta R,\quad (\rm IW).
\end{align} 

In unmagnetized ideal fluids, the IW parametric instability excites daughter IW modes with arbitrary large wavenumbers, including large $m$. In the presence of a strong toroidal magnetic field $\bB_0$, magnetic tension arrests excitation of fluid oscillations with $|m|\gg 2\Omega/\omA$. The parametric instability is then free to pump daughter modes with any large $k_r$ and $k_\theta$, and $|m|\lesssim 2\Omega/\omA$. This behavior is observed in our simulations (which have modest $2\Omega/\omA\sim 3$): secondary small-scale IWs from the parametric instability of the linear TI modes are observed in the poloidal plane (Figure~\ref{fig:NL}, right column). By contrast, our simulations do not show evidence for parametric instabilities of MW (and there are no analytical studies of such instabilities to our knowledge). The observed saturation of the MW mode of TI is consistent with the two nonlinear processes described by timescales $t_0$ and $t_{\rm cas}$.

\subsection{Saturation amplitudes}

The linear phase ends when the shortest nonlinear timescale becomes comparable to the TI timescale,
\begin{align}
    \min\{t_{0},t_{\mathrm{cas}},t_{\rm pi}\}\sim \gamma^{-1}.
\end{align}
Note that $t_0 \lesssim t_{\rm cas}$ for MW and $t_{\rm pi}<t_{\rm cas}<t_{0}$ for IW. This can be verified using $|\bB_1|/\sqrt{4\pi\rho}\sim (2\Omega/\omA) |\bu_1|$ for MW and $|\bB_1|/\sqrt{4\pi\rho}\sim (\omA/2\Omega)|\bu_1|$ for IW, where the eigenmode amplitudes can be approximated by their horizontal components (e.g. $|u_{\phi,1}|\sim |u_{\theta,1}|\sim |\bu_{1}|$).

Consider first the saturation of MW. Assuming no parametric instability, we find the saturated amplitude from $t_0^{-1}\sim \gamma$:
\begin{align}
    \label{eq:MW_sat}
    |\bu_1|\sim \frac{\omA}{\Omega}\frac{B_0}{\sqrt{4\pi\rho}},\quad |\bB_{1}|\sim B_0,\quad (\rm MW).
\end{align}
Saturation thus occurs with disruption of the background field by $\delta B\sim|\bB_1|\sim B_0$ and large horizontal displacements of the magnetic loops $\xi_h \sim |\bu_{1}/\omega|\sim R$. 

Next, consider the saturation of IW. In this case, the parametric instability is the fastest nonlinear process. Saturation occurs when $t_{\rm pi}\sim \gamma^{-1}$, which gives
\begin{align}
    \label{eq:IW_sat}
    |\bu_1|\sim \frac{\omA}{\Omega}\frac{1}{\keta R}\frac{B_0}{\sqrt{4\pi\rho}},\quad |\bB_1|\sim\left(\frac{\omA}{\Omega}\right)^2 \frac{B_0}{\keta R},\quad (\rm IW).
\end{align}
The IW amplification by the TI thus stops at relatively small amplitudes, $\delta B\ll B_0$, and with small displacements of magnetic loops $\xi_{h}\sim |\bu_1/\omega|\sim R (\omA/\Omega)^2(\keta R)^{-1}\ll R$. The power input from the TI, $\dot E\sim \rho|\bu_1|^2/2t_{\rm pi}$, is channeled into short-wavelength daughter waves through the parametric instabilities, driving a non-local forward cascade. Meanwhile, the background field remains nearly unaffected because the energy in the saturated IWs is small. Following initial saturation, one can expect a nonlinear, quasi-steady evolution of the system as the background field is gradually dissipated on the much longer timescale $\sim B_0^2/8\pi\dot E\sim\gamma^{-1}(\Omega/\omA)^2(\keta R)^2\sim t_{0}.$ 

These theoretical scalings for the saturated amplitudes agree well with our measurements of $\delta u_\theta$ and $\delta B$ across the parameter scans presented in Figure~\ref{fig:Scaling}. Individual simulations in Figure~\ref{fig:NL} also broadly support the predictions. The MW saturate with large amplitudes and destroy the axisymmetric background field on magnetostrophic timescales, consistent with the prediction $t_{0}\sim \gamma^{-1}$. Simulations do not show short-wavelength perturbations developing on top of the MW, supporting our assumption that  parametric instabilities are negligible.\footnote{Parametric instability of MW with growth timescale $t_{\rm pi}^{-1}\sim |k_\phi u_{\phi,1}|\kTI R$ would predict saturation with $|B_{\phi ,1}|/B_0\sim (\kTI R)^{-1}$, which is ruled out by the numerical scan.} TI saturation in the IW mode occurs at smaller amplitudes, consistent with the estimated timescale for parametric instabilities. The daughter modes created by the parametric instability are observed as small-scale variations in the poloidal plane, in particular in the $\theta$-direction, with $k_\theta R\gg 1$; these secondary modes are distinct from the primary TI modes that have $k_\theta R\sim 1$. After the saturation of IW, the  decay of the background field gradually develops. The decay takes a much longer time, consistent with $t_0\gg t_{\rm pi}$ estimated above.

\subsection{Radial field regeneration}

It has long been debated whether the TI can successfully close the TSD loop by regenerating axisymmetric radial fields. Radial fields with $m=0$ clearly cannot be generated by shearing the radial component of $|m|=1$ eigenmodes by differential rotation \citep{zahn2007magnetic}. However, they can be generated through nonlinear induction \citep{fuller2019slowing,ji2023magnetohydrodynamic}. Estimating the radial field generation rate in \Eq~(\ref{eq:B0_evol}) gives:
\begin{align}
\label{eq:dtBR0}
    |\partial_t B_{r,0}|&=|\left[\nabla\times(\langle\bu_1\times\bB_1\rangle_0)\right]_r|\nonumber\\
    &\sim\frac{| B_{\phi,0}|}{t_{0}}\frac{| B_{r,1}|}{| B_{\phi,1}|}\sim \frac{B_0}{t_{0}\kTI R}.
\end{align}
Evaluating this nonlinear term using $\bu_1$ and $\bB_1$ of the eigenmodes (MW or IW) shows that the constant prefactor in the above estimate is $\sim 1$ (Appendix~\ref{ap:NL}).

In the MW case of TI, generation during the linear and early saturation phases on the magnetostrophic timescale $\gamma^{-1}\sim t_{0}$ results in an axisymmetric radial field $| B_{r,0}|\sim\gamma^{-1}|\partial_t B_{r,0}| $, which yields
\begin{align}
\label{eq:BR_sat_MW}
    | B_{r,0}|\sim\frac{B_0}{\kTI R},\quad\rm
    (MW).
\end{align}
This estimate is consistent with the $\kkappa$ and $\knu$ simulations, which sustain $|B_{r,0}|\sim B_0/\kTI R$ deep into the saturation phase (Figure~\ref{fig:NL}). We also note that $|B_{r,0}|$ estimated in \Eq~(\ref{eq:BR_sat_MW}) is comparable to the critical value for suppressing TI by the magnetic tension of the radial field. This marginal suppression regime is commonly invoked for the TSD \citep{spruit2002dynamo,fuller2019slowing}.

In the IW case, axisymmetric radial field generation is less efficient because of the longer nonlinear evolution timescale for axisymmetric fields $t_{0}\gg \gamma^{-1}$. Using \Eq~(\ref{eq:dtBR0}), the magnitude of the radial field  generated over the linear and early saturation phase ($\sim \gamma^{-1}|\partial_t B_{r,0}|$) is 
\begin{align}
\label{eq:BR_sat_IW}
    |B_{r,0}|\sim  \frac{| B_{\phi,0}|}{\gamma t_{0}\kTI R}\sim \frac{B_0}{(\kTI R)^3} \left(\frac{\omA}{\Omega}\right)^2,\quad \rm
    (IW).
\end{align}
It is smaller than for MW (\Eq~\ref{eq:BR_sat_MW})  by a factor of $(\kTI R)^{-2}(\omA/\Omega)^2$. Significant magnetic diffusion (required for the $\kTI=\keta$ mode of TI) can reduce $B_{r,0}$.

The scaling of $ | B_{r,0}|\sim \overline B_r$ with rotation from the numerical scan shown in Figure~\ref{fig:Scaling} agrees well with the estimate for MW (\Eq~\ref{eq:BR_sat_MW}), while the estimate for IW (\Eq~\ref{eq:BR_sat_IW}) is slightly steeper than the data, which may be due to magnetic diffusion effects neglected in \Eq~(\ref{eq:BR_sat_IW}). We caution that the estimates are formally valid only during the linear instability phase and early saturation phase while nonlinearities are weak. Additionally, in the full TSD scenario with differential rotation and a turbulent steady state, the generation of $ B_{r,0}$ may be influenced by the shear flow and contributions of higher azimuthal modes.

\section{Discussion}
\label{sec:Conc}

We have carried out global simulations of a toroidal field in a rotating, stably stratified star to examine the linear and nonlinear phases of the TI. The simulations confirm the new linear theory developed in \citetalias{Skoutnev_2024}, which extends and corrects the commonly invoked marginal stability analysis in \cite{spruit1999differential}. In particular, the simulations demonstrate the key role of three diffusivities: thermal conductivity $\kappa$, viscosity $\nu$, and magnetic diffusivity $\eta$. The toroidal field configuration remains in stable equilibrium in simulations with equal diffusivities, and becomes unstable when diffusivities are significantly different. As predicted, the TI develops via overstability of either MW or IW, depending on the thermal and magnetic Prandtl numbers of the fluid, $ Pr=\nu/\kappa$ and $Pm=\nu/\eta$. The radial wavenumbers of the TI are confirmed to be $\kkappa$ and $\knu$ for the MW branch (enabled by thermal diffusion and viscosity, respectively), and $\keta$ for the IW branch (enabled by magnetic diffusion).

We then used our simulation results to examine TI saturation. We also developed and tested an analytical framework for saturation of MW and IW branches. The saturation level of overstable TI wave modes (all of which have $|m|=1$) is determined by nonlinear processes that can stop their exponential growth. We identified three such processes: (1) weakening of the axisymmetric toroidal field ($m=0$) that drives the instability, (2) energy transfer to $|m|=2$ modes, initiating the forward cascade to higher $m$, and (3) parametric instability of the amplified waves that transfers energy directly to much smaller scales.

The MW saturation observed in the simulations is well explained by disruption of the axisymmetric field followed by a forward cascade. At saturation, perturbations reach large amplitudes and displace toroidal loops across the polar region $(\delta B/ B_0\sim \xi_h/R\sim1)$. Dissipation of the background field energy occurs on the magnetostrophic timescale of the instability, consistent with the saturation model in \cite{spruit2002dynamo}.

For IW, the parametric instabilities have the shortest timescale and dominate the saturation dynamics. The amplified IW eigenmodes saturate  with small amplitudes and small displacements of magnetic loops $(\delta B/ B_0\sim \xi_h/R\ll1)$, channeling their energy directly to small scales. The background field energy then slowly dissipates over timescales longer than the instability timescale, resembling the picture that \cite{fuller2019slowing} proposed for MWs. 

Our global numerical model demonstrates the TI mechanism and helps improve understanding of the TSD. Saturation of MW modes with $\delta B\sim B$ rules out the boosted TSD scenario based on small-amplitude saturation \citep{fuller2019slowing} and supports the TSD model of \cite{spruit2002dynamo}. The simulations also shed some light on other aspects of the TSD loop. The turbulence is driven by the TI at the poles and may not encompass the entire star, failing to reach the equatorial regions containing a significant fraction of the AM and rotational energy. A global picture of the TSD may require secondary transport processes such as hydrodynamic instabilities and meridional circulation \citep{zahn1992circulation} that can couple the polar dynamo region to the equatorial AM reservoir. Another aspect is the regeneration of poloidal fields. Our simulations of a uniformly rotating star show that axisymmetric poloidal fields are efficiently generated by nonlinear effects during the instability of MW modes. On the other hand, poloidal field regeneration is highly inefficient for the $\keta$ mode of the TI, suggesting that the TSD may not operate via the IW branch. However, we caution that self-consistent closure of the dynamo loop requires including differential rotation, which is actively being explored in global models \citep{petitdemange2023spin,barrere2023numerical,barrere2026coexisting}. 

Our results favoring the \cite{spruit2002dynamo} model of the TSD imply that the TSD alone cannot account for the efficient AM transport inferred in evolving low-mass stars \citep{cantiello2014angular,fuller2019slowing}. The TSD is further limited by its inability to operate in compositional layers outside the stellar cores, where the TI is suppressed\footnote{Compositional layers can have $Pm<1$ and be unstable to IW. However, the minimum values of $Pm={\cal O}(10^{-1})$ are not substantially below unity, so IW instability is unlikely to be robust.} \citep{skoutnev2025zones}. Explaining efficient AM transport in stars remains an open challenge, particularly across compositionally stratified layers (for a review, see \citealt{aerts2019angular}). One partial solution is efficient AM transport in regions with magnetic fields locked into stable configurations dubbed ``magnetic webs" \citep{skoutnev2025magnetic}. Stars may host multiple disconnected magnetic webs deposited by successive convective episodes. Differential rotation would then exist only in gaps between webs, and the TSD or other instability-based mechanisms must bridge the gaps. 1D stellar models show that a single web in the core (deposited during the main sequence), combined with the TSD prescription of \cite{spruit2002dynamo} outside the web, can explain slow core rotation on the early red giant branch until the growing core buries the web \citep{skoutnev2025magnetic}. The scenario of multiple webs with the TSD transport between them requires more exploration. 

The TSD is also invoked to transport AM and magnetize nascent neutron stars formed in supernova or binary mergers. Current models typically adopt the TSD prescription of \cite{fuller2019slowing} as an upper bound on AM transport. Adopting the less efficient TSD prescription of \cite{spruit2002dynamo} would increase the predicted lifetime of hypermassive neutron stars before their collapse into black holes \citep{margalit2022angular}, affect modeling of multi-messenger signatures from long-lived merger remnants \citep{ng2026spinning}, and reduce the predicted magnetization of surface layers during supernova fallback \citep{barrere2022}. We also note that the updated linear theory in \citetalias{Skoutnev_2024} and the results on TI saturation presented here can improve our understanding of the physics and resolution requirements needed to successfully capture the TSD in  simulations of binary neutron star mergers \citep{reboul2025tayler,ng2026spinning}.

\begin{acknowledgments}
This work is supported by NSF grant AST-2408199. A.M.B. is also supported by NASA grants 21-ATP21-0056 and 80NSSC24K1229, and Simons Foundation grant 446228. This work was facilitated by MPPC grant PHY-2206609.
\end{acknowledgments}

%

\vspace{5mm}





\appendix

\section{Parameter scan of the Rotation Rate}
\label{ap:ParameterScan}

We perform a parameter scan to determine the scaling of the TI saturation amplitudes with the rotation rate. The scan varies the rotation parameter, $2\Omega/\omA$, while fixing the super-criticality, maintaining a constant ratio between the Alfv\'en frequency $\omA$ and its minimum (or maximum) critical value for instability onset. This preserves the ratios of relevant linear terms in the MHD equations as the rotation rate is varied. Below, we describe how super-criticality is fixed for each TI mode, which prescribes the dependence of $N_0/2\Omega,Pr, Pm,$ and $ Lu$ on the rotation parameter. 

For the $\keta$ mode, the scan is performed in the limit $\eta\gg\nu=\kappa$, so the thermal Prandtl number is always unity, $Pr=1$. In the rotating regime $2\Omega/\omA>1$, the instability condition for negligible viscosity is $\omA>2\Omega Pm^{1/2}$ (\Eq~\ref{eq:keta_cond}), which defines the super-criticality parameter $C_{\eta}=\omA/(2\Omega Pm^{1/2})$. Thus, fixing $C_{\eta}$ requires $Pm=C_{\eta}^{-2}(\omA/2\Omega)^2$. We additionally fix $N_0/2\Omega=const$ to maintain the ordering $N_0>2\Omega>\omA$. The remaining condition is that stratification  does not suppress instability, $k_\eta>k_N$, which requires $Lu=Lu_0(2\Omega/\omA)$ with a sufficiently large $Lu_0$. The complete set of parameters used in the scan is: $C_\eta=4$, $Pm=C_\eta^{-2}\min\{1,(\omA/2\Omega)^2\}$, $N_0/2\Omega=2\cdot\max\{1,\omA/2\Omega\}$, $Lu=1500 \max\{1,2\Omega/\omA\}$, and $Pr=1$. Here, the use of the maximum and minimum functions ensures a continuous transition to parameter values in the nonrotating regime where $2\Omega/\omA<1$. 

For the $\knu$ mode, the scan is performed in the limit  $\nu\gg\eta=\kappa$, implying $Pr=Pm$. The required parameter scalings are analogous to those for the $\keta$ mode. The instability condition (negligible magnetic diffusivity) requires $\omA>2\Omega/Pm^{1/2}$ (\Eq~\ref{eq:knu_cond}) and the super-criticality parameter is $C_{\nu}=(\omA/2\Omega) Pm^{1/2}$. The complete parameter set used in the scan is: $C_\nu=4$, $Pm=C_\nu^{2}\max\{1,(2\Omega/\omA)^2\}$, $N_0/2\Omega=2\cdot\max\{1,\omA/2\Omega\}$, $Lu=1500 Pm \max\{1,2\Omega/\omA\}$, and $Pr=Pm$. 

For the $\kkappa$ mode, the scan is performed in the limit $\kappa\gg\nu=\eta$, so $Pm=1$. This mode is more complicated since it has both a minimum and maximum critical value of $\omA$ (\Eq~\ref{eq:kkappa_cond}), which define two criticality parameters:
\begin{align}
    C_{\kappa,1}&=\frac{\omA}{\om}\left(\frac{N}{\om} \frac{\eta}{\kappa}\right)^{-1/2}\left(\frac{\kappa k_\theta^2}{\om}\right)^{-1/4},\quad C_{\kappa,2}=\frac{\om}{\omA}\left(\frac{N}{\om} \right)^{1/2} \left(\frac{\kappa k_\theta^2}{\om}\right)^{1/4}.    
\end{align}
For fixed $N_0/2\Omega$, the required scalings are  $\eta/\kappa=(C_{\kappa,1}C_{\kappa,2})^{-2}$ and $Lu=(C_{\kappa,1}/C_{\kappa,2})^2(2\Omega/\omA)^3(N_0/2\Omega)^2$.  The complete set of parameters used in the scan is: $C_{\kappa,1}=5$, $C_{\kappa,2}=2$,  $N_0/2\Omega=16\cdot\max(1,\omA/2\Omega)$, $Lu=1600 \max(1,(2\Omega/\omA)^3)$, $Pr=1/100$, and $Pm=1$.

The scalings adopted above cause all three characteristic wavenumbers of TI to increase with the rotation parameter: $\keta R,\knu R,\kkappa R\propto 2\Omega/\omA$. The resulting shorter radial wavelengths, smaller diffusivities, and slower TI growth combine to increase the computational cost of exploring faster rotation rates with global simulations. Our resolution constraints thus limited the parameter scans to maximum values of  $2\Omega/\omA=3.5,6,$ and $3.5$ for the $\keta,\knu$ and $\kkappa$ modes, respectively. Faster rotation rates can be probed in local simulations where rescaling of the equations reduces some of these costs \citep{fraser2026}.

We identify the saturation point of the TI differently for the MW and IW modes due to their different saturation dynamics. The MW modes typically saturate once the perturbation amplitudes reach a well-defined peak and begin to gradually decay. We note that in the $\kkappa$ simulations, saturation does not occur simultaneously in the domain because the linear stratification profile $N^2\propto r$  produces a spatial variation in the local TI growth rate and wavenumber. Saturation begins near the core, where stratification is weakest, and then progresses outward. We extract the saturated amplitudes when $\delta B_\theta$ peaks, which has a better defined peak than $\delta B$, since $\delta B\propto |\bB-\bB_0|$ tends to grow or level off as the background field is destroyed. This coincides with saturation throughout most of the domain and visible decline of the background field amplitude $\overline B_\phi/B_0$. For the IW mode, saturation occurs more gradually because the background field is weakly affected and the TI remains active. The velocity perturbation amplitudes do not exhibit a peak but instead slowly level out, whereas the magnetic perturbations $\delta B\propto |\bB-\bB_0|$ keep growing as the background field gradually dissipates. We therefore determine the saturation point by where the velocity amplitudes of $\delta u_\theta$ begin to level off. Uncertainties (the error bars in Figure~\ref{fig:Scaling}) are estimated by taking the standard deviation of the amplitudes over a time range spanning two dynamical timescales before and after the determined saturation point.

\section{Nonlinear coupling of azimuthal modes}
\label{ap:NL}
The TI amplifies only eigenmodes with $|m|=1$. Its saturation depends on nonlinear coupling between the amplified eigenmodes and other azimuthal modes $|m|\neq 1$. The coupling arises through the nonlinear induction, Lorentz, and advection terms in the induction and momentum equations. These terms can be evaluated analytically for the ``Malkus" field, $\bB_M\propto r\sin \theta \, \hat\phi$ \citep{malkus1967hydromagnetic}, which approximates the background field near the polar axis where $|\cos\theta|\approx1$. Below, we evaluate the nonlinear terms and show that the induction term, in particular, has a magnitude comparable to the upper bound estimate derived in Section~\ref{sec:Theory}.

We first obtain the TI eigenmodes from linear theory and then use them to compute the nonlinear terms. In the remainder of this Appendix, we focus on the polar region of a star and adopt cylindrical coordinates $(s,\phi,z)$. The background field is taken to be the Malkus field $\bB_M=B_0(s/s_0)\hat\phi$, where $s_0$ is a reference radius. A gas pressure gradient supports the field in equilibrium, requiring no background buoyancy field $\vartheta_0=0$. The linearized MHD Boussinesq equations for perturbations $\bu,\bb,\vartheta$ about the background state $\bu_0=0,\bB_0=\bB_M,\vartheta_0=0$ are 
\begin{align}
    \frac{\partial \bu}{\partial t}&=-\frac{\nabla P'}{\rho }+\vartheta \hat z -2\Omega\hat z\times\bu+\frac{(\nabla\times\bB_0)\times\bb}{4\pi\rho}+\frac{(\nabla\times\bb)\times\bB_0}{4\pi \rho }+\nu\nabla^2\bu,\\
    \frac{\partial \bb}{\partial t}&=\nabla\times(\bu\times \bB_0)+\eta\nabla^2\bb,
    \\
    \frac{\partial \vartheta}{\partial t}&=-N_0^2u_z+\kappa \nabla^2\vartheta.
\end{align}
where the stable stratification is approximated as locally constant.

Assuming perturbations of the form $f(s,\phi,z,t)=\Re[\tilde f(s,z)\exp(im\phi-i\omega t)]$, the linearized equations can be reduced to a single equation for the velocity field:
\begin{align}
    I(\omega)\tilde\bu&=-i\omega_\eta\frac{\nabla P''}{\rho}+S(\omega)\tilde u_z\hat z-C(\omega)\hat z\times\tilde\bu,\\
    I(\omega)\equiv\omega_\nu\omega_\eta  -m^2&\omA^2,\quad 
    S(\omega)\equiv\frac{N_0^2\omega_\eta}{\omega_\kappa},\quad 
    C(\omega)\equiv2i\Omega  \omega_\eta+2im\omA^2,
\end{align}
where  $P''$ denotes the total pressure, and $\omega_\nu\equiv \omega+i\nu k^2$, $\omega_\eta\equiv \omega+i\eta k^2$, $\omega_\kappa\equiv \omega+i\kappa k^2$.  Here, we approximate all diffusive terms with their WKB expression ($\nabla^2\approx k^2$) appropriate for the characteristic short radial scales of TI modes, where $k$ is the total wavenumber defined below. A standard form of the above equation is the Poincar\'e equation. It is obtained by taking the curl of the velocity equation twice and then examining the vertical component:
\begin{align}
    &\left[\nabla^2-\frac{1}{\lambda^2}\partial_z^2\right]u_z=0,\quad
    \frac{1}{\lambda^2}\equiv\frac{C(\omega)^2+I(\omega)S(\omega)}{-I(\omega)^2+I(\omega)S(\omega)}.
\end{align}
Once a domain and boundary conditions are specified, this equation defines an eigenvalue problem with eigenvalues satisfying $0\leq\lambda^2<1$ \citep{greenspan1968theory}. In the WKB limit, the above equation recovers the familiar TI dispersion relation, $D(\omega)=0,$ given by Equation~(20) in \citetalias{Skoutnev_2024}, which can be seen by identifying $\nabla^2 u_z\approx -k^2 u_z$, $\partial_z^2u_z\approx -k_z^2u_z$, and $\lambda^2\approx k_z^2/k^2\approx 1$ (also see \cite{fraser2026}).

For concreteness, let us now consider a cylindrical domain of height $H$ and radius $s_0$ with periodic boundary conditions in the vertical direction and impenetrable boundary conditions on the side walls, $u_s(s_0)=0$. Perturbations can be taken in the simpler form $f(s,z,\phi,t)=\Re[\tilde f(s)\exp(ik_zz+im\phi-i\omega t)]$. The Poincar\'e equation for each mode then depends only on the radial coordinate and can be transformed into Bessel's equation of order $\alpha=m$ by introducing the dimensionless coordinate  $x=k_zs\sqrt{1/\lambda^2-1}\equiv k_h s$. The quantity $k_h$ is interpreted as the horizontal wavenumber using the WKB relation $\lambda^2\approx k_z^2/k^2$ with $k^2=k_h^2+k_z^2$.  The non-divergent solutions are composed of Bessel functions of the first kind: 
\begin{align}
    u_z&=v_{\rm A}\sum_{\lambda,k_z,m}\frac{k_h}{k_z}J_m(x)\,\Re[a_{\lambda,k_z,m}e^{i(k_zz+m\phi-\omega_rt)}]e^{\gamma t},
\end{align}
where $a_{\lambda,k_z,m}$ is a dimensionless complex amplitude satisfying $|a_{\lambda,k_z,m}|\ll1$ during the linear phase of the TI. The radial and azimuthal velocity components are found from the incompressibility equation and the vertical component of the curl of the momentum equation: 
\begin{align}
    u_s&=v_{\rm A}\sum_{\lambda,k_z,m}\Re\left[ia_{\lambda,k_z,m}\left(\partial_x J_m(x)-\frac{m}{\lambda' x}J_m(x)\right)e^{i(k_zz+m\phi-\omega_rt)}\right]e^{\gamma t},\\
    u_\phi&=v_{\rm A}\sum_{\lambda,k_z,m}\Re\left[a_{\lambda,k_z,m}\left(\frac{1}{\lambda'}\partial_x J_m(x)-\frac{m}{x}J_m(x)\right)e^{i(k_zz+m\phi-\omega_rt)}\right]e^{\gamma t}.
\end{align}
where $\lambda'=iI(\omega)/C(\omega)$ is a complex number with magnitude $|\lambda'|={\cal O}(1)$ for MW and IW modes. The magnetic field components follow from the linearized induction equation, e.g. for a single mode $\bb=\Re[-m\omA\bu \sqrt{4\pi\rho}/\omega_\eta]$. The eigenvalue value spectrum needed to specify the eigenmodes requires evaluating the boundary conditions. However, we do not derive it here because the exact eigenvalue spectrum is not required for the calculation of the nonlinear terms below (for a derivation, see \citealt{zhang2017theory}). 

We evaluate the nonlinear terms when $\bu$ and $\bb$ are composed of only a single $m=1$  mode with $\lambda\approx1$ and $k_z= \kTI$. This mode is representative of the fastest growing Tayler modes and contributes to the $|m|=1$ fields, $\bu_1$ and $\bb_1$, defined in Section~\ref{sec:Theory}. The vertical velocity component becomes $u_{1,z}=a_1v_{\rm A}(k_h/k_z)J_1(x)\cos(k_zz+\phi-\omega_rt)e^{\gamma t}$, where we define $a_1=a_{\lambda,k_z,1}$ and choose $a_1$ to be real. The other field components follow from the expressions above. Evaluating the nonlinear induction term, we find: 
\begin{align}
\left[\nabla\times(\bu\times\bb)\right]_s=&0,\\
\label{eq:dtBphi_eigenmodes_phi}
    \left[\nabla\times(\bu\times\bb)\right]_\phi=&-k_ha_1^2v_AB_0\frac{(\gamma+\eta k^2) \omA}{|\omega_\eta|^2}\,\partial_x\left[\Re\left[\left(\frac{\partial_x J_1(x)}{\lambda'^*}-\frac{J_1(x)}{x}\right)\left(\partial_x J_1(x)-\frac{J_1(x)}{\lambda' x}\right)\right] \right]e^{2\gamma t},\\
\label{eq:dtBphi_eigenmodes_z}
    \left[\nabla\times(\bu\times\bb)\right]_z=&-\frac{k_h^2}{k_z}a_1^2v_AB_0\frac{(\gamma+\eta k^2) \omA}{|\omega_\eta|^2}\frac{1}{x}\,\partial_x \left[xJ_1(x)\left(\partial_x J_1(x)-\frac{J_1(x)}{ x}\Re\left[\frac{1}{\lambda'}\right]\right) \right]e^{2\gamma t}.
\end{align}

These results demonstrate that a single eigenmode nonlinearly couples to $m=0$ modes through the induction equation. The strength of the coupling has a similar magnitude to the  estimates in \Eqs~(\ref{eq:t_nl_0}) and (\ref{eq:dtBR0}). For instance, the magnitude of the $\phi$ component (\Eq~\ref{eq:dtBphi_eigenmodes_phi}) is $\sim k_h a_1^2v_{\rm A}B_0(\gamma+\eta k^2)\omA/|\omega_\eta|^2$ since $|J_1(x)|\sim |\partial_xJ_1(x)|\sim {\cal O}(1)$ for the most unstable Tayler modes with $x\sim 1$. This agrees with \Eq~(\ref{eq:t_nl_0}) after identifying $k_\phi\sim k_h$, $|u_{\phi,1}|\sim |a_1v_A|$, $|B_{\phi,1}|\sim |a_1B_0\omA/\omega_\eta|$, and noting that $(\gamma+\eta k^2)/|\omega_\eta|={\cal O}(1)$ for TI modes. The magnitude of the vertical component in \Eq~(\ref{eq:dtBphi_eigenmodes_z}) is smaller by a factor of $k_h/k_z\sim (\kTI s_0)^{-1}$, consistent with the estimate in \Eq~(\ref{eq:dtBR0}).

We note that a single eigenmode for the Malkus field does not nonlinearly couple to $|m|=2$ modes through the induction equation. However, using the notation of Section~\ref{sec:Theory}, one can show that the nonlinear coupling in the momentum equation generates a $\bu_2$ velocity field that subsequently couples to $\bB_2$  through the ${\cal L}_2$ term in the induction equation.  In particular, we find that for MW, magnetostrophic balance $|2\Omega\hat z\times\bu_2|\sim|\bB_1\cdot \nabla \bB_1/4\pi\rho|$ gives $|u_{\phi,2}|\lesssim|B_{\phi,1}^2k_\phi|/4\pi\rho\Omega$, and for IW, inertia-Coriolis balance $|2\Omega\hat z\times\bu_2|\sim|\bu_1\cdot \nabla \bu_1|$ gives $|u_{\phi,2}|\lesssim|u_{\phi,1}^2k_\phi|/\Omega$. The resulting magnitude of $|\bB_2| \sim|\! \int dt\,\nabla\times(\bu_2\times\bB_0)|\lesssim |u_{\phi,2}k_\phi|B_0/2\gamma$ is comparable to the estimate in \Eq~(\ref{eq:B_2_mag}) for both MW and IW.


\bibliography{refs}{}

@article{burns2020dedalus,
  title={Dedalus: A flexible framework for numerical simulations with spectral methods},
  author={Burns, Keaton J and Vasil, Geoffrey M and Oishi, Jeffrey S and Lecoanet, Daniel and Brown, Benjamin P},
  journal={Physical Review Research},
  volume={2},
  number={2},
  pages={023068},
  year={2020},
  publisher={APS}
}

@article{reboul2025tayler,
  title={Tayler-Spruit dynamo in binary neutron star merger remnants},
  author={Reboul-Salze, Alexis and Barr{\`e}re, Paul and Kiuchi, Kenta and Guilet, J{\'e}r{\^o}me and Raynaud, Rapha{\"e}l and Fujibayashi, Sho and Shibata, Masaru},
  journal={A \& A},
  volume={699},
  pages={A4},
  year={2025},
  publisher={EDP Sciences}
}

@article{Skoutnev_2024,
year = {2024},
month = {oct},
publisher = {The American Astronomical Society},
volume = {974},
number = {2},
pages = {290},
author = {Valentin A. Skoutnev and Andrei M. Beloborodov},
title = {Tayler Instability Revisited},
journal = {ApJ},
}

@article{skoutnev2025zones,
  title={Zones of Tayler instability in stars},
  author={Skoutnev, Valentin A and Beloborodov, Andrei M},
  journal={ApJ},
  volume={988},
  number={2},
  pages={195},
  year={2025},
  publisher={The American Astronomical Society}
}

@article{spiegel1982magnetic,
  title={Magnetic buoyancy and the Boussinesq approximation},
  author={Spiegel, EA and Weiss, NO},
  journal={GAFD},
  volume={22},
  number={3-4},
  pages={219--234},
  year={1982},
  publisher={Taylor \& Francis}
}

@article{gottlieb2024she,
  title={She’s Got Her Mother’s Hair: Unveiling the Origin of Black Hole Magnetic Fields through Stellar to Collapsar Simulations},
  author={Gottlieb, Ore and Renzo, Mathieu and Metzger, Brian D and Goldberg, Jared A and Cantiello, Matteo},
  journal={ApJL},
  volume={976},
  number={1},
  pages={L13},
  year={2024},
  publisher={The American Astronomical Society}
}

@article{barrere2026coexisting,
	author = {Barr{\`e}re, Paul and Reboul-Salze, A. and Eggenberger, P. and Deheuvels, S. and Rodríguez, C. and Marchand, M.},
	title = {Coexisting Tayler instability-driven dynamos in radiative zones: New dynamo solution and its impacts on stellar physics},
	journal = {A\&A},
	year = 2026,
	volume = 713,
	pages = "A62",
}

@article{meduri2025gravity,
  title={Gravity’s role in taming the Tayler instability in red giant cores},
  author={Meduri, Domenico G and Arlt, Rainer and Bonanno, Alfio and Licciardello, Giovanni},
  journal={A\&A},
  volume={702},
  pages={A44},
  year={2025},
  publisher={EDP Sciences}
}

@article{malkus1967hydromagnetic,
  title={Hydromagnetic planetary waves},
  author={Malkus, Willem VR},
  journal={JFM},
  volume={28},
  number={4},
  pages={793--802},
  year={1967},
  publisher={Cambridge University Press}
}

@article{gusakov2026beyond,
  title={Beyond the Tayler instability: A new global instability of toroidal magnetic fields in stars},
  author={Gusakov, Mikhail E and Becerra, Laura and Kantor, Elena M and Reisenegger, Andreas and Valdivia, Juan Alejandro},
  journal={Phys. Rev. D.},
  volume={114},
  number={2},
  pages={023019},
  year={2026},
  publisher={APS}
}

@article{barker2010internal,
  title={On internal wave breaking and tidal dissipation near the centre of a solar-type star},
  author={Barker, Adrian J and Ogilvie, Gordon I},
  journal={MNRAS},
  volume={404},
  number={4},
  pages={1849--1868},
  year={2010},
  publisher={Blackwell Publishing Ltd Oxford, UK}
}

@article{skoutnev2025magnetic,
  title={Magnetic Webs in Stellar Radiative Zones},
  author={Skoutnev, Valentin A and Beloborodov, Andrei M},
  journal={ApJL},
  volume={989},
  number={1},
  pages={L4},
  year={2025},
  publisher={The American Astronomical Society}
}

@article{barrere2023numerical,
  title={Numerical simulations of the Tayler--Spruit dynamo in proto-magnetars},
  author={Barr{\`e}re, Paul and Guilet, J{\'e}r{\^o}me and Raynaud, Rapha{\"e}l and Reboul-Salze, Alexis},
  journal={MNRAS: Letters},
  volume={526},
  number={1},
  pages={L88--L93},
  year={2023},
  publisher={Oxford University Press}
}

@article{ma2019angular,
  title={Angular momentum transport in massive stars and natal neutron star rotation rates},
  author={Ma, Linhao and Fuller, Jim},
  journal={MNRAS},
  volume={488},
  number={3},
  pages={4338--4355},
  year={2019},
  publisher={Oxford University Press}
}

@article{spruit1999differential,
  title={Differential rotation and magnetic fields in stellar interiors},
  author={Spruit, HC},
  journal={A\&A},
  volume={349},
  pages={189},
  year={1999}
}

@article{rincon2019dynamo,
  title={Dynamo theories},
  author={Rincon, Fran{\c{c}}ois},
  journal={JPlPh},
  volume={85},
  number={4},
  pages={205850401},
  year={2019},
  publisher={Cambridge University Press}
}

@book{greenspan1968theory,
  title={The theory of rotating fluids},
  author={Greenspan, Harvey P},
  year={1968},
  publisher={Cambridge University Press}
}

@book{zhang2017theory,
  title={Theory and modeling of rotating fluids: convection, inertial waves and precession},
  author={Zhang, Keke and Liao, Xinhao},
  year={2017},
  publisher={Cambridge University Press}
}

@ARTICLE{Kerswell1999,
       author = {{Kerswell}, R.~R.},
        title = "{Secondary instabilities in rapidly rotating fluids: inertial wave breakdown}",
      journal = {JFM},
         year = 1999,
        month = mar,
       volume = {382},
       number = {1},
        pages = {283-306}
}

@ARTICLE{Lifshitz1996,
       author = {{Lifschitz}, Alexander and {Fabijonas}, Bruce},
        title = "{A new class of instabilities of rotating fluids}",
      journal = {Physics of Fluids},
         year = 1996,
        month = aug,
       volume = {8},
       number = {8},
        pages = {2239-2241}
}

@ARTICLE{Miyazaki1998,
       author = {{Miyazaki}, Takeshi and {Lifschitz}, Alexander},
        title = "{Three-Dimensional Instabilities of Standing Waves in Rotating Fluids}",
      journal = {Journal of the Physical Society of Japan},
         year = 1998,
        month = apr,
       volume = {67},
       number = {4},
        pages = {1226}
}

@ARTICLE{Mora2021,
       author = {{Mora}, Daniel Odens and {Monsalve}, Eduardo and {Brunet}, Maxime and {Dauxois}, Thierry and {Cortet}, Pierre-Philippe},
        title = "{Three-dimensionality of the triadic resonance instability of a plane inertial wave}",
      journal = {Phys. Rev. Fluids},
         year = 2021,
        month = jul,
       volume = {6},
       number = {7},
          eid = {074801},
        pages = {074801}
}

@article{AM2024,
  title={Short-wavelength instabilities in a finite-amplitude plane inertial wave},
  author={Abhiram, DS and Mathur, Manikandan},
  journal={JFM},
  volume={982},
  pages={A22},
  year={2024},
  publisher={Cambridge University Press}
}

@article{ascher1997implicit,
  title={Implicit-explicit Runge-Kutta methods for time-dependent partial differential equations},
  author={Ascher, Uri M and Ruuth, Steven J and Spiteri, Raymond J},
  journal={Applied Numerical Mathematics},
  volume={25},
  number={2-3},
  pages={151--167},
  year={1997},
  publisher={Elsevier}
}

@article{fraser2026,
  title={},
  author={Fraser, Adrian and Skoutnev, Valentin and Lecoanet, Daniel},
  journal={In preparation.},
  volume={},
  pages={},
  year={2026},
  publisher={}
}

@article{fuller2019most,
  title={Most black holes are born very slowly rotating},
  author={Fuller, Jim and Ma, Linhao},
  journal={ApJL},
  volume={881},
  number={1},
  pages={L1},
  year={2019},
  publisher={The American Astronomical Society}
}

@article{ng2026spinning,
  title={Spinning Down Neutron-star Merger Remnants with the Tayler--Spruit Dynamo: Global Simulations Reveal the Formation of Massive Disks and Neutron-rich Ejecta},
  author={Ng, Harry Ho-Yin and Most, Elias R},
  journal={ApJL},
  volume={1010},
  number={2},
  pages={L42},
  year={2026},
  publisher={The American Astronomical Society}
}

@article{vasil2019tensor,
  title={Tensor calculus in spherical coordinates using Jacobi polynomials. Part-I: mathematical analysis and derivations},
  author={Vasil, Geoffrey M and Lecoanet, Daniel and Burns, Keaton J and Oishi, Jeffrey S and Brown, Benjamin P},
  journal={JCP: X},
  volume={3},
  pages={100013},
  year={2019},
  publisher={Elsevier}
}

@article{lecoanet2019tensor,
  title={Tensor calculus in spherical coordinates using Jacobi polynomials. Part-II: implementation and examples},
  author={Lecoanet, Daniel and Vasil, Geoffrey M and Burns, Keaton J and Brown, Benjamin P and Oishi, Jeffrey S},
  journal={JCP: X},
  volume={3},
  pages={100012},
  year={2019},
  publisher={Elsevier}
}

@article{monteiro2023global,
  title={Global simulations of Tayler instability in stellar interiors: a long-time multistage evolution of the magnetic field},
  author={Monteiro, G and Guerrero, G and Del Sordo, F and Bonanno, A and Smolarkiewicz, PK},
  journal={MNRAS},
  volume={521},
  number={1},
  pages={1415--1428},
  year={2023},
  publisher={Oxford University Press}
}

@article{bonanno2013rotational,
  title={Rotational suppression of the Tayler instability in stellar radiation zones},
  author={Bonanno, Alfio and Urpin, Vadim},
  journal={MNRAS},
  volume={431},
  number={4},
  pages={3663--3669},
  year={2013},
  publisher={Oxford University Press}
}

@article{spiegel1960boussinesq,
  title={On the Boussinesq approximation for a compressible fluid.},
  author={Spiegel, Edward A and Veronis, G},
  journal={ApJ},
  volume={131},
  pages={442},
  year={1960}
}

@article{petitdemange2023spin,
  title={Spin-down by dynamo action in simulated radiative stellar layers},
  author={Petitdemange, Ludovic and Marcotte, Florence and Gissinger, Christophe},
  journal={Sci},
  volume={379},
  number={6629},
  pages={300--303},
  year={2023},
  publisher={American Association for the Advancement of Science}
}

@article{fuller2019slowing,
  title={Slowing the spins of stellar cores},
  author={Fuller, Jim and Piro, Anthony L and Jermyn, Adam S},
  journal={MNRAS},
  volume={485},
  number={3},
  pages={3661--3680},
  year={2019},
  publisher={Oxford University Press}
}

@article{barrere2022,
  title={A new scenario for magnetar formation: Tayler-Spruit dynamo in a proto-neutron star spun up by fallback},
  author={Barr{\`e}re, P and Guilet, J{\'e}r{\^o}me and Reboul-Salze, Alexis and Raynaud, Rapha{\"e}l and Janka, H-T},
  journal={A\&A},
  volume={668},
  pages={A79},
  year={2022},
  publisher={EDP Sciences}
}

@article{skoutnev2026stability,
  title={Stability of propagating plane inertial waves in rotating fluids},
  author={Skoutnev, Valentin and Astoul, Aur{\'e}lie and Barker, Adrian J},
  journal={Phys. Rev. Fluids},
  volume={11},
  number={2},
  pages={024802},
  year={2026},
  publisher={APS}
}

@article{zahn1992circulation,
  title={Circulation and turbulence in rotating stars},
  author={Zahn, J-P},
  journal={A\&A},
  volume={265},
  pages={115--132},
  year={1992}
}

@article{margalit2022angular,
  title={Angular-momentum Transport in Proto-neutron Stars and the Fate of Neutron Star Merger Remnants},
  author={Margalit, Ben and Jermyn, Adam S and Metzger, Brian D and Roberts, Luke F and Quataert, Eliot},
  journal={ApJ},
  volume={939},
  number={1},
  pages={51},
  year={2022},
  publisher={IOP Publishing}
}

@article{tayler1973adiabatic,
  title={The Adiabatic Stability of Stars Containing Magnetic Fields--I: T OROIDAL F IELDS},
  author={Tayler, RJ},
  journal={MNRAS},
  volume={161},
  number={4},
  pages={365--380},
  year={1973},
  publisher={Oxford University Press Oxford, UK}
}

@article{guerrero2019global,
  title={Global simulations of Tayler instability in stellar interiors: the stabilizing effect of gravity},
  author={Guerrero, G and Del Sordo, F and Bonanno, A and Smolarkiewicz, PK},
  journal={MNRAS},
  volume={490},
  number={3},
  pages={4281--4291},
  year={2019},
  publisher={Oxford University Press}
}

@article{bonanno2012breakdown,
  title={Breakdown of chiral symmetry during saturation of the Tayler instability},
  author={Bonanno, Alfio and Brandenburg, Axel and Del Sordo, Fabio and Mitra, Dhrubaditya},
  journal={PhRvE},
  volume={86},
  number={1},
  pages={016313},
  year={2012},
  publisher={APS}
}

@article{ji2023magnetohydrodynamic,
  title={Magnetohydrodynamic simulations of the Tayler instability in rotating stellar interiors},
  author={Ji, Suoqing and Fuller, Jim and Lecoanet, Daniel},
  journal={MNRAS},
  volume={521},
  number={4},
  pages={5372--5383},
  year={2023},
  publisher={Oxford University Press}
}

@article{weber2015tayler,
  title={The Tayler instability at low magnetic Prandtl numbers: between chiral symmetry breaking and helicity oscillations},
  author={Weber, Norbert and Galindo, Vladimir and Stefani, Frank and Weier, Tom},
  journal={NJPh},
  volume={17},
  number={11},
  pages={113013},
  year={2015},
  publisher={IOP Publishing}
}

@article{zahn2007magnetic,
  title={On magnetic instabilities and dynamo action in stellar radiation zones},
  author={Zahn, J-P and Brun, AS and Mathis, St{\'e}phane},
  journal={A\&A},
  volume={474},
  number={1},
  pages={145--154},
  year={2007},
  publisher={EDP Sciences}
}

@article{pitts1985adiabatic,
  title={The adiabatic stability of stars containing magnetic fields--VI. The influence of rotation},
  author={Pitts, E and Tayler, RJ},
  journal={MNRAS},
  volume={216},
  number={2},
  pages={139--154},
  year={1985},
  publisher={Oxford University Press Oxford, UK}
}

@article{cantiello2014angular,
  title={Angular momentum transport within evolved low-mass stars},
  author={Cantiello, Matteo and Mankovich, Christopher and Bildsten, Lars and Christensen-Dalsgaard, J{\o}rgen and Paxton, Bill},
  journal={ApJ},
  volume={788},
  number={1},
  pages={93},
  year={2014},
  publisher={IOP Publishing}
}

@article{aerts2019angular,
  title={Angular momentum transport in stellar interiors},
  author={Aerts, Conny and Mathis, St{\'e}phane and Rogers, Tamara M},
  journal={ARA\&A},
  volume={57},
  pages={35--78},
  year={2019},
  publisher={Annual Reviews}
}

@article{spruit2002dynamo,
  title={Dynamo action by differential rotation in a stably stratified stellar interior},
  author={Spruit, HC},
  journal={A\&A},
  volume={381},
  number={3},
  pages={923--932},
  year={2002},
  publisher={EDP Sciences}
}
\bibliographystyle{aasjournal}



\end{document}